\documentclass[12pt,a4paper]{article}
\usepackage{amsmath,epsfig,graphicx,amssymb}
\usepackage[latin1]{inputenc}
\usepackage[margin=1in]{geometry}
\usepackage{amsmath}
\usepackage{amsfonts}
\usepackage{amssymb}
\usepackage{makeidx}
\usepackage{graphicx}
\usepackage{setspace}
\newcommand{\beq}{\begin{equation}}
\newcommand{\eeq}{\end{equation}}
\newcommand{\ben}{\begin{eqnarray}}
\newcommand{\een}{\end{eqnarray}}
\newcommand{\bes}{\begin{subequations}}
\newcommand{\ees}{\end{subequations}}
\newcommand{\bFig}{\begin{figure}}
\newcommand{\eFig}{\end{figure}}

\date{}
\begin{document}

\title{Entanglement in Classical Optics}
\author{Partha Ghose\footnote{partha.ghose@gmail.com} \\
Centre for Astroparticle Physics and Space Science (CAPSS),\\Bose Institute, \\ Block EN, Sector V, Salt Lake, Kolkata 700 091, India, \\and\\ Anirban Mukherjee\footnote{mukherjee.anirban.anirban@gmail.com},\\Indian Institute of Science Education \& Research,\\Mohanpur Campus, West Bengal 741252.}
\maketitle
{\bf Keywords}: classical light, polarization, entanglement, nonseparability, noncontextuality, Pancharatnam phase, information processing
\begin{abstract}
The emerging field of entanglement or nonseparability in classical optics is reviewed, and its similarities with and differences from quantum entanglement clearly pointed out through a recapitulation of Hilbert spaces in general, the special restrictions on Hilbert spaces imposed in quantum mechanics and the role of Hilbert spaces in classical polarization optics. The production of Bell-like states in classical polarization optics is discussed, and new theorems are proved to discriminate between separable and nonseparable states in classical wave optics where no discreteness is involved. The influence of the Pancharatnam phase on a classical Bell-like state is deived. Finally, to what extent classical polarization optics can be used to simulate quantum information processing tasks is also discussed. This should be of great practical importance because coherence and entanglement are robust in classical optics but not in quantum systems.  

\end{abstract}

\tableofcontents
\newpage
\section{Introduction}
Entanglement is a concept that was introduced into physics by Schr\"{o}dinger in 1935 \cite{schr} although essentially the same notion was also the central point of the famous EPR paper \cite{epr}. Coupled with the measurement problem in quantum mechanics, this concept has been believed to lie at the heart of quantum mechanics and the difficulties of its interpretation, including its nonlocal character. Let us consider the maximally entangled state
\beq
\vert \psi\rangle = \frac{1}{\sqrt{2}}\left[\vert H\rangle_L \vert V\rangle_R + \vert V\rangle_L \vert H\rangle_R \right]
\eeq
where, for example, $H$ and $V$ could denote horizontal and vertical polarization states of photons and $L$ and $R$ denote left and right moving modes. Then, a measurement on any one of the photons will show it to be either in the $H$ or $V$ state with equal probability. Once a particular state (say $\vert H\rangle_L$) is detected on the left, the other photon is predicted to be certainly in the state $\vert V\rangle_R$ according to the first term. This is because the other possibility represented by the second term disappears, and the entangled state $\vert \psi\rangle$ is projected to, i.e. collapses to a product state. The other possibility, namely that the first term disappears and only the second term is left is also possible. It is this `projective measurement' that implies a nonlocal influence of one photon on its distant and noninteracting partner. Hence, without the additional postulate of projective measurement in quantum mechanics, entanglement {\em per se} does not imply nonlocality. Entanglement {\em per se} is an inevitable consequence of the Hilbert space structure of a theory and nothing else.

Hilbert spaces, however, also occur in areas of mathematics and physics that have nothing to do with quantum mechanics. For example, classical electrodynamics has a Hilbert space structure. Once this is recognized, it becomes at once clear that entanglement is possible in classical electrodynamics \cite{spreeuw}, \cite{ghose1} without implying the special features associated with quantum projective measurement such as nonlocality. To appreciate this important point, we will first review the basic features of Hilbert spaces in Section 2. In Section 3 we will hightlight the additional assumptions made in quantum mechanics which are not shared by other theories. The basis of the Hilbert space structure of classical polarization optics will be explained in Section 4. In Section 5 we will demonstrate how Bell-like states of classical light can be generated and how such states violate noncontextuality, a requirement of innocuous classical realism. In Section 6, we will introduce nonseparable two-mode classical states and their implications for local realism. The influence of Bell-like classical states on the Pancharatnam phase will be demonstrated in Section 7. In Section 8 we will briefly outline how some key quantum information processing protocols can be performed using classical polarization optics. In the final Section 9, we will indicate a potentially useful area of application of the Hilbert space structure of classical field theories, namely human cognition.

\section{Salient Features of Hilbert Spaces}

Let $\cal{H}$ be a complex linear vector space of dimension $n$ spanned by the complete set of basis vectors $\{\vert x_i)\}$  ($i = 1,2,...,n$) and its dual $\cal{H}^*$ by the complete set of basis vectors $\{( y_j\vert\}$ so that every $\vert X) \in \cal{H}$ can be expressed as $\vert X) = \sum_{i=1}^n c_i \vert x_i)$ and every $( Y\vert \in \cal{H}^*$ as $( Y\vert = \sum_{i=1}^n d_i ( y_i \vert$ where $c_i$ and $d_i$ are complex coefficients. (We will use the notation $\vert X)$ and $(Y\vert$ to denote nonquantum states in Hilbert space.) The inner product on $\cal{H}$ must be specified so as to have the following properties:
\begin{enumerate}
\item
$(X\vert X)\geq 0$\,\, $\forall \vert X) \in\cal{H}$ and $(X\vert X) = 0$ iff $\vert X) = 0$;
\item
$( a Y\vert b X) = a^* b (Y\vert X)$\,\, $\forall \vert X) \in \cal{H}$ and $\forall (Y\vert \in \cal{H}^*$; and
\item
$(Y\vert X)  = \overline{( X\vert Y)}$\,\, $\forall \vert X) \in \cal{H}$ and $\forall (Y\vert \in \cal{H}^*$.
\end{enumerate}
The length of a vector $\vert X)$ is usually denoted by $\vert \vert X \vert \vert$ and the inner product $(Y\vert X) = \vert \vert X \vert \vert\, \vert \vert Y \vert \vert\, {\rm cos}\theta$ in the real case. Hence, $\vert\vert X\vert\vert = \sqrt{( X\vert X)} = \sum_i \vert c_i\vert^2$. The basis vectors are usually chosen to be orthonormal, i.e. $( x_j\vert x_i) = \delta_{ij}$. Being endowed with an inner product and an induced length or norm, $\cal{H}$ is called a Hilbert space. For example, the space $\mathbb{C}^n$ of n-tuples of complex numbers $z = (z_1, z_2, ...,z_n)$ and $z^\prime = (z_1^\prime, z_2^\prime, ...,z_n^\prime)$ is a Hilbert space under the inner product defined by $\langle z, z^\prime\rangle = \sum_{k=1}^n  z_k \bar{z}_k^\prime$. The square integrable functions on $L^2$ also form a Hilbert space. For any $f$ and $g$ in $L^2$ one defines the inner product $\langle f, g\rangle = \int_{\Omega}f(x) \overline{g(x)} dx$. Hence, $\langle f, f\rangle < \infty$. 

{\flushleft{{\em Superposition principle}}}

The linearity of Hilbert spaces allows vectors to be superposed to form other vectors in the same space. Thus, for $\vert X), \vert Y) \in {\cal{H}}$, $\vert X) + \vert Y) \in {\cal{H}}$. An example is the interference of two coherent classical waves at a point $x$. Let $F_1(x, t) = a e^{i(\phi_1(x) - \omega t)}$ and $F_2(x, t) = b e^{i(\phi_2 (x) -\omega t)}$ be two waves at $x$ at time $t$. Then, the sum of the two waves is $G(x, t) = F_1(x, t) + F_2(x, t)$ and the intensity at $x$ is given by $I(x) = \langle G(x, t), G(x, t)\rangle = \int G(x, t) \overline{G(x, t)} dt = a^2 + b^2 + 2 ab\, {\rm cos} (\phi_1 - \phi_2)$. The second term gives rise to the familiar interference pattern. 

The direct sum of two Hilbert spaces ${\cal{H}}_1$ and ${\cal{H}}_2$ (${\cal{H}}_1 \cap {\cal{H}}_2 = \emptyset$) of dimensions $n_1$ and $n_2$ is the set ${\cal{H}}_1 \oplus {\cal{H}}_2$ of pairs of vectors ($\vert h_1\rangle, \vert h_2\rangle)$ in ${\cal{H}}_1$ and ${\cal{H}}_2$ with the operations 
\begin{enumerate}
\item
$(\vert h_1), \vert h_2)) + (\vert h_1^\prime), \vert h_2^\prime)) = (\vert h_1) + \vert h_1^\prime), \vert h_2) + \vert h_2^\prime))$
\item
$c(\vert h_1), \vert h_2)) = (c\vert h_1), c\vert h_2))$ 
\end{enumerate}
Its dimension is ($n_1 + n_2$).

{\flushleft{{\em Projection operators}}}
  
One can define projection operators $\pi_i = \vert x_i)( x_i\vert$ for every basis vector $\vert x_i)$ which are hermitian ($\pi_i^\dagger = \pi_i$) and idempotent ($\pi_i. \pi_i = \pi_i$). They have the important property that $\pi_i (1 - \pi_i) = 0$, i.e. a projector and its complement are orthogonal. Thus, $\pi_i\vert X) = \sum_j c_j \vert x_i)( x_i\vert \vert x_j) = c_i \vert x_i)$, and hence $c_i = ( x_i\vert\pi_i\vert X)$ and $\vert c_i\vert^2 = ( X\vert\pi_i\vert X)$. Thus, $\vert c_i\vert$ is a measure of the membership of $\vert x_i)$ in $\vert X)$.

{\flushleft{{\em Linear transformations and choice of basis}}}

An important characteristic of vector spaces is the complete freedom to choose any set of basis vectors related by unitary transformations. Thus, for example, one can express $\vert X) = \sum_i c_i \vert x_i)$ or as $\vert X) = \sum_i c_i^\prime \vert x_i^\prime)$ provided $\vert x_i^\prime) = \sum_j S_{ij} \vert x_j)$ with $S$ a unitary matrix, i.e. $S^\dagger S = S S^\dagger = 1$ so that $S^\dagger = S^{-1}$. Operators $\cal{O}$ (represented by $n \times n$ matrices in the old basis) acting on $\cal{H}$ then transform to the new basis as ${\cal{O^\prime}} = S {\cal{O}}S^{-1}$ which is a similarity transformation.

{\flushleft{{\em Operators on Hilbert space and groups}}}

Bounded operators on a Hilbert space can form {\em non-Abelian} Lie groups and generate their algebras. A simple example is the rotation group $SO(3)$ in Euclidean space $\mathbb{R}^3$. As is well known, rotations along the three Cartesian axes do not commute, and this non-commutation has nothing to do with quantum mechanical incompatibility and uncertainty. Rather, it reflects certain symmetry properties of Euclidean space, and is correctly described by the non-Abelian property of the group $SO(3)$. $SO(3)$ has a universal covering group $SU(2)$, the group of all $2\times 2$ unitary matrices with complex elements and determinant $1$. Although the group $SU(2)$ is used in quantum physics to describe particles with spin $\frac{1}{2}\hbar$ (fermions), the group itself is independent of quantum mechanics, and can be used to describe any two-valued objects. These two examples should suffice to make the point clear that non-Abelian groups of operators act naturally on Hilbert spaces, and their non-commutative structure has nothing to do with quantum physics. This property of Hilbert spaces should be borne in mind in modeling human cognitive processes.

{\flushleft{{\em Tensor products}}} 

Apart from the direct sum ${\cal{H}}_1 \oplus {\cal{H}}_2$ of two Hilbert spaces, one can also form tensor products ${\cal{H}} = {\cal{H}}_1 \otimes {\cal{H}}_2$ whose dimension is $(n_1 \times n_2)$. In fact, there is a mathematical theorem which states that {\em every pair of vector spaces has a tensor product} \cite{stern}. Let $\{\vert e_i) \} (1 \leq i \leq n_1)$ and $\{\vert f_j)\} (1 \leq j \leq n_2)$ be the sets of basis vectors of ${\cal{H}}_1$ and ${\cal{H}}_2$ respectively. Then, the set $\{\vert e_i) \otimes \vert f_j)\} ( \leq i < n_1, 1 \leq j < n_2)$ forms the basis of ${\cal{H}}$. Here, $\{\vert e_i) \otimes \vert f_j)\}$ stands for the $(n_1 \times n_2)$ matrix whose $i$th row consists of the ordinary products $\vert e_i) \vert f_j) (1 \leq j \leq n_2)$. Hence it has $n_1$ rows and $n_2$ columns. A typical element of ${\cal{H}}_1 \otimes {\cal{H}}_2$ would be $\sum_{i,j}c_{ij} (\vert e_i) \otimes \vert f_j))$. Since this tensor product space is also a Hilbert space, one can define an inner product on it by

$$( (( e_i\vert \otimes ( f_i\vert)(\vert e_j) \otimes \vert f_j)) ) = ( e_i\vert e_j) ( f_i\vert f_j)$$ 

Now consider two $2$-dimensional Hilbert spaces ${\cal{H}}_A$ and ${\cal{H}}_B$ with basis vectors $\vert e_A) = (\vert h)_A, \vert v)_A)$ and $\vert e_B) = (\vert h)_B, \vert v)_B)$. One can construct the following four states from them:

\ben
\vert \Phi^+) &=& \frac{1}{\sqrt{2}} [\vert h)_A\otimes \vert h)_B + \vert v)_A\otimes \vert v)_B ]\label{b1}\\
\vert \Phi^-) &=& \frac{1}{\sqrt{2}} [\vert h)_A\otimes \vert h)_B - \vert v)_A\otimes \vert v)_B ]\label{b2}\\
\vert \Psi^+) &=& \frac{1}{\sqrt{2}} [\vert h)_A\otimes \vert v)_B + \vert v)_A\otimes \vert h)_B ]\label{b3}\\
\vert \Psi^-) &=& \frac{1}{\sqrt{2}} [\vert h)_A\otimes \vert v)_B - \vert v)_A\otimes \vert h)_B ]\label{b4}
\een
These states are of fundamental significance. When two states $\vert A) = \sum_i c_i \vert e_i)$ and $\vert B) =\sum_j d_j \vert f_j)$ are independent of one another, one can construct a product state $\vert X) = \vert A) \otimes \vert B) = \sum_{i,j} c_i d_j \vert e_i) \otimes \vert f_j)$, and the state $\vert X)$ is said to be `factorizable' into $\vert A)$ and $\vert B)$ which retain their identity, independence and separability. However, an inspection of (\ref{b1}, \ref{b2}, \ref{b3}, \ref{b4}) shows that the basis states $(\vert h)_A, \vert v)_A)$ and $(\vert h)_B, \vert v)_B)$ cannot be factored out of the four states $\vert \Phi^+), \vert \Phi^-), \vert \Psi^+), \vert \Psi^-)$. Hence, none of these four states is factorizable. Such states may be called `entangled states' because the states from which they are constructed lose their identity, independence and separability in such states. The four states above form a special set of basis states which are maximally entangled, and are analogs of the Bell states in quantum mechanics \cite{bell}, i.e. they are Bell-like states. Notice that so far, Hilbert spaces have not been associated with any physical systems, classical or quantum.
 
\section{Hilbert Spaces in Quantum Mechanics}

In this section we list the additional restrictions that are imposed on Hilbert spaces to get quantum mechanics which is so very different from Newtonian classical physics. 
\begin{enumerate}
\item
First, in order to have a probabilistic interpretation, quantum state vectors are all normalised (i.e. of unit norm). Hence, the set of all pure states corresponds to the unit sphere in Hilbert space, with the additional requirement that all vectors that differ only by a complex scalar factor (a phase factor) are identified with the same state. Thus, quantum mechanics operates on coset spaces and Grassmanian manifolds \cite{chatur}.

As a consequence, in quantum mechanics, unless a state is an eigenstate of some observable, a state does not {\em possess} physical properties {\em before} measurement. An example is a single-photon state like $\vert X\rangle = c_1\vert H\rangle + c_2 \vert V\rangle$ with $\vert c_1\vert^2 + \vert c_2\vert^2 = 1$. It cannot be said to possess any polarization $H$ or $V$. When measured, however, this state is projected to either $\vert H\rangle$ with probability $\vert c_1\vert^2$ or $\vert V\rangle$ with probability $\vert c_2\vert^2$. In classical electrodynamics which is deterministic a state $\vert X) = c_1\vert H) + c_2\vert V)$ of light with classical amplitudes $c_1$ and $c_2$ {\em always} possesses a definite polarization. This realism of the polarization states in classical electrodynamics, however, holds only for product states like $(c_1 \vert x) + c_2 \vert y))\otimes \vert w)$ but not for a superposition of product states like $c_1 \vert u) \otimes \vert x) + c_2 \vert v) \otimes \vert y)$. 
  
\item
Second, a linear and unitary equation of motion, the Schr\"{o}dinger equation, is postulated that specifies the time evolution of states in Hilbert space. 
\item
Third, all observables are represented by hermitian operators ${\cal{O}}^\dagger = {\cal{O}}$ on Hilbert space. Let $\vert \Psi\rangle = \sum_i c_i \vert \psi_i\rangle$ be a pure state, and let $\rho = \vert \Psi\rangle\langle \Psi\vert$ be the density operator which has the properties $\rho^2 = \rho$ and ${\rm Tr} \rho^2 = 1$. Then, the expectation value of ${\cal{O}}$ is given by $\bar{\cal{O}} = {\rm Tr} \rho \,{\cal{O}}$. The results of observations are obtained by projective measurements $M_i = \vert \psi_i\rangle\langle \psi_i\vert$ acting on the state $\vert \Psi\rangle: \hat{\rho} = \sum_i M_i \rho M_i^\dagger$. It is clear that ${\rm Tr}\hat{\rho}^2 < 1$. $\hat{\rho}$ is called the reduced density operator, and represents mixed states. It is straightforward to show that $\hat{\rho}_{ij} = \vert c_i\vert^2 \delta _{ij} $. This projection is additional to the linear and unitary time evolution. This is the measurement postulate. Its ad hoc and non-unitary character has spawned a plethora of interpretations of quantum mechanics \cite{int}.
\item
Fourth, every pair of canonical dynamical variables ($ p_i, q_j$) is postulated to be represented by hermitian operators ($\hat{p}_i, \hat{q}_j$) with the commutation rules $[\hat{p}_i, \hat{q}_j] = - i\hbar \delta_{ij}$ resulting in the famous Heisenberg uncertainty relations $\Delta q\, \Delta p \geq \hbar/2$. This is the canonical quantization postulate. These commutation relations vanish in the formal limit $\hbar \rightarrow 0$, and hence do not arise from a non-abelian character of the operators and are not intrinsic to Hilbert spaces.
\item
Fifth, entangled states in quantum mechanics are extremely {\em fragile}, i.e. the entanglement disappears very rapidly and the system decoheres and breaks into pieces when exposed to an environment \cite{zurek}. Such is not the case with classical electrodynamics in which coherence and entanglement/nonseparability are robust.

\end{enumerate}

We will now pass on to a brief description of the mathematical structure of classical polarization optics.

\section{Hilbert Space and Classical Polarization\\ Optics}

The Hilbert space structure of classical electrodynamics was first explicilty used by Spreeuw \cite{spreeuw} and independently demonstrated by Ghose and Samal \cite{ghose1}. The complete description of an ordinary state in classical electrodynamics involves the direct product of two disjoint Hilbert spaces, namely a space ${\cal{H}}_{path}$ of square integrable functions and a two-dimensional space of polarization states ${\cal{H}}_{pol}$. Hence, a state of unit intensity can be written as $\frac{1}{\sqrt{\vert A\vert^2}}\vert A)\otimes \vert \lambda)\in {\cal{H}}_{path}\otimes {\cal{H}}_{pol}$ where $A({\bf r},t)$ are solutions of the scalar wave equation
and $\vert \lambda) \in {\cal{H}}_{pol}$ is the vector 
\beq
\vert \lambda) = e^{i\phi}\left(\begin{array}{c}\cos\theta \\ e^{i\chi}\sin\theta\end{array}\right)
\eeq 
of the transverse polarizations $\lambda_1$ and $\lambda_2$. This can also be written as the Jones vector \[\vert J) = \frac{1}{\sqrt{(J\vert J)}}\left(\begin{array}{c}
 E_x\\ E_y
\end{array} \right) \] with $E_x = A_0 \hat{e}_x {\rm exp (i\phi_x)}$ and $E_y = A_0 \hat{e}_y {\rm exp (i\phi_y)}$ the complex transverse electric fields, $\hat{e}_x$ and $\hat{e}_y$ the unit polarization vectors, and $(J\vert J) = \vert E_x\vert^2 + \vert E_y\vert^2 =  A_0^2$ the intensity $I_0$.
Polarization-path entanglement in classical polarization optics is an inevitable consequence of this mathematical structure. 

{\flushleft{\em Entanglement and Classical Polarization States}}

That classical optical fields like thermal light are necessarily entangled (in the sense of non-factorizability or nonseparability of the two disjoint vector spaces corresponding to the spatial and polarization degrees of freedom) has been recently shown by Xiao-Feng Qian and Eberly \cite{eberly}. They have shown that the degree of polarization of a light field corresponds to the degree of separability of these two spaces. This leads to a natural measure of the degree of polarization applicable to any optical field, whether beamlike or not. It turns out that only homogeneously polarized light beams correspond to fully separable or factored states. Interestingly, an ideal thermal light field is shown to be a Bell state that violates a Bell inequality, showing that typical Bell correlations are not exclusive to quantum states. 

{\flushleft{\em Cylindrically Polarized Light}}

Nonseparable cylindrically polarized laser beams have been extensively studied and used since 1993 \cite{oron, radial, holleczek, gabriel}. 
The scalar field distributions of the $TEM_{01(x)}$ and $TEM_{01(y)}$ Laguerre-Gaussian modes are given by
\ben
E_x(r, \theta) &=& E_0 \sqrt{\rho} \exp (-\rho/2)\cos \theta,\\
E_y (r, \theta) &=& E_0 \sqrt{\rho}\exp (-\rho/2)\sin \theta, 
\een
where $r$ and $\theta$ are the cylindrical coordinates, $E_0$ the magnitude of the electric field, $\rho = 2r^2/w^2$ with $w$ as the waist of the Gaussian beam \cite{oron}. A coherent summation of such modes with orthogonal polarizations leads to either azimuthally or radially polarized modes
\ben
E_\theta (r, \theta) &=& \hat{y}E_x(r, \theta) - \hat{x}E_y (r, \theta) = \hat{\theta}E_0 \sqrt{\rho} \exp (-\rho/2),\\
E_r (r, \theta) &=& \hat{x}E_x(r, \theta) + \hat{y}E_y (r, \theta) = \hat{r}E_0 \sqrt{\rho} \exp (-\rho/2).
\een
These states are evidently polarization-position entangled and robust. Fig. 1 is a representation of azimuthally polarized light.

\begin{figure}
\centering
{\includegraphics[scale=0.6]{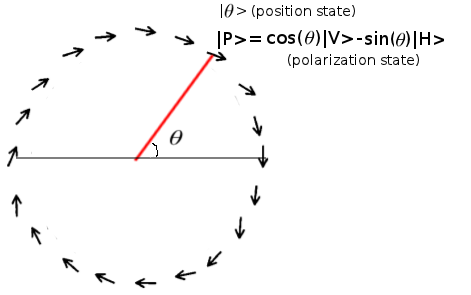}}
\caption{\label{Figure 1}{\footnotesize Variation of polarization with space of an emerging azimuthally polarized beam.}}
\end{figure}

{\flushleft{\em Mueller Matrices and Entanglement}}

Recently, it has been shown that non-quantum entanglement plays an essential role in resolving the basic issue of the choice of the appropriate subset of $4\times 4$ real matrices that should be accepted as Mueller matrices in classical optics \cite{simon}. While Jones matrices are adequate to describe fully polarized light, Mueller matrices are required to describe the transformation of all forms of light passing optical elements. Any light beam, polarized or unpolarized, can be described by the Stokes vector $\vec{S}$. After it passes a linear optical element, it is transformed by the Mueller matrix $M: \vec{S^\prime} = M \vec{S}$ where $M$ is a $4\times 4$ real matrix. Let $\Omega^{pol}$ be the state space of all real vectors $\vec{S} \in \mathbb{R}^4$. It turns out that although all real $4\times 4$ matrices map $\Omega^{pol}$ into itself, not all of them are physical Mueller matrices. To be considered as physical Mueller matrices, they must satisfy a stronger positivity criterion that follows from nonseparability or entanglement of path and polarization of inhomogeneous light fields.

{\flushleft{\em Bell-like Inequality Criterion for Entangled Classical Light}}

Borges et al \cite{borges,gabriel} have proposed an inequality criterion in analogy with Bell's inequality for the separability of the spin and orbit degrees of freedom of a laser beam and shown that this inequality is violated by classical optical states for which the spin-orbit degrees are nonseparable. However, the deeper significance of such a criterion is not clear.  

{\flushleft{\em Classical Vortex Beams with Topological Singularities}}

That light can be twisted like a corkscrew around its axis of travel was first pointed out by Nye and Berry \cite{nye}. This twisting results in the cancellation of light amplitudes on the axis. Such light is called an optical vortex. When incident on a flat surface, an optical vortex looks like a ring of light with a dark hole in the centre. This has led to the discovery of classical optical beams with topological singularities and charges (the number of twists in one wave length) that have diverse applications such as in optical tweezers, and to the study of the coherence properties of vortex beams. Such beams have Schmidt decomposition and share many properties with quantum optical systems.  Using the Wigner function, Chowdhury et al have shown that the coherence properties of such beams imply violations of Bell inequalities for continuous variables \cite{agarwal}.

\vskip 0.1in

Notwithstanding all this, it must be pointed out that classical and quantum entanglement have significantly different {\em implications}, and hence the appropriate nomenclature to be used in the case of classical light has been a matter of some debate. Some prefer to use `nonseparability' \cite{borges}, some `structural inseparability' \cite{gabriel} and some `non-quantum entanglement' \cite{simon} for classical light. We prefer to use the term `classical entanglement' though nonseparability will also be used equivalently at times. This is to emphasize the fact that non-factorizability is not an exclusive feature of quantum mechanics; it is intrinsic to classical optics as well. Contrary to expectations, classical wave optics turns out to be more analogous to quantum mechanics than is Newtonian mechanics. 

Classical entanglement can be used to simulate the manipulations necessary for quantum information processing except those depending on quantum nonlocality. This was first shown by Spreeuw \cite{spreeuw}. The fact that classical entanglement is robust is an added advantage. 

Another very important aspect of classical entanglement is its bearing on the concept of noncontextuality and realism. It has recently been shown that the polarization and spatial modes of classically entangled light are {\em contextual} variables \cite{ghose2}, which is a real surprise in classical physics. We proceed now to see how this comes about.

\section{Bell-like States in Classical Optics and Noncontextuality}

\begin{figure}[!ht]
{\includegraphics[scale=0.6]{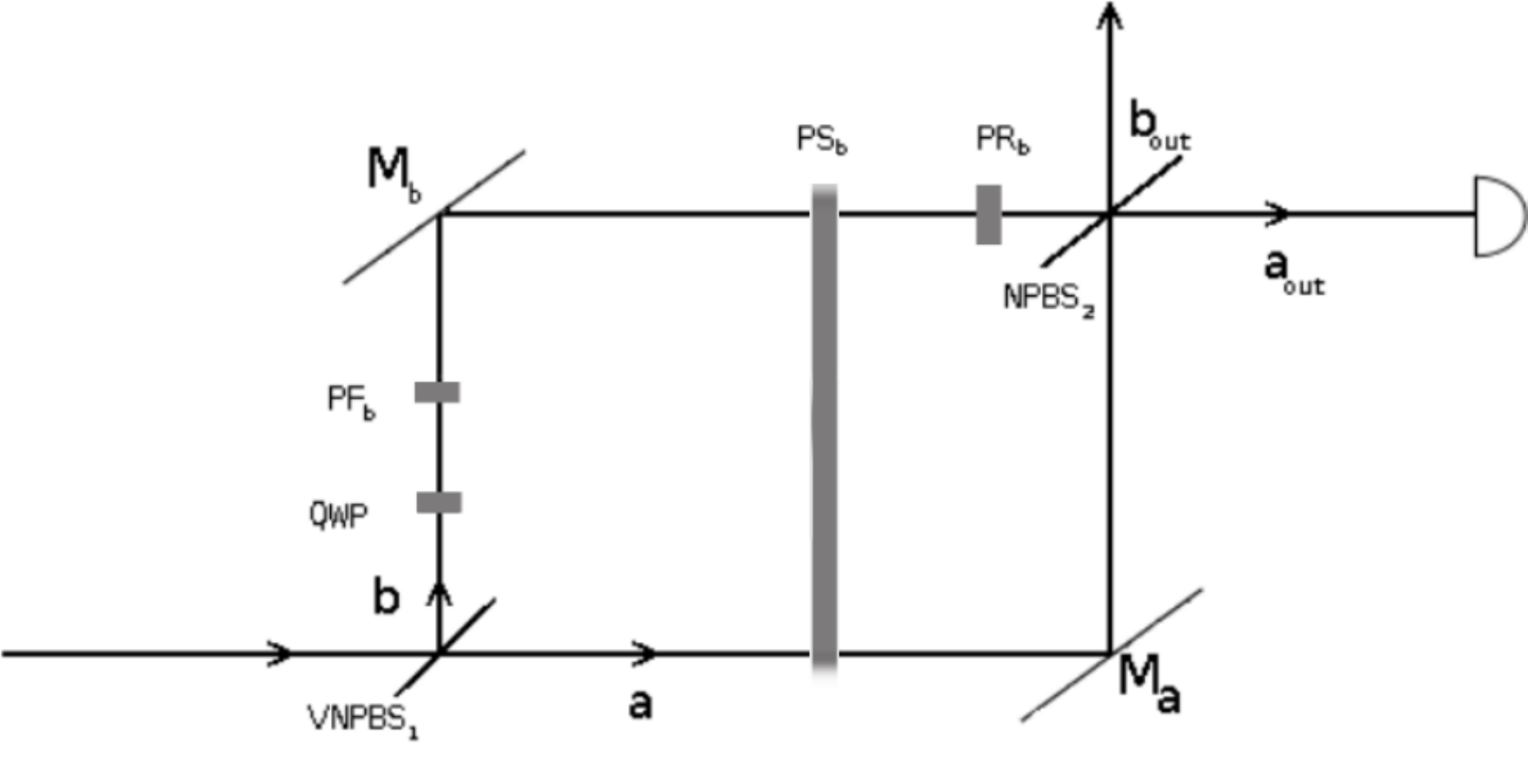}}
\caption{\label{Figure 2}{\footnotesize Schematic diagram: An H polarized classical light beam passes through a Mach-Zehnder interferometer. $PF_b$ is a polarization flipper that converts $V$ to $H$, $PS_b$ is a phase-shifter, and $PR_b$ is a polarization rotator.}}
\end{figure}

Let us consider an incoming vertically polarized monochromatic classical light beam of intensity $I_0 = \vert A \vert^2$ incident on a 50-50 lossless non-polarizing beam splitter $NPBS_1$ (Fig. 2). The transmitted beam $\vert a)$ and the reflected beam $\vert b)$ span a two-dimensional Hilbert space $\hat{H}_{path}=\lbrace\vert a),\vert b)\rbrace$, and the polarization states of the two outgoing light beams span a disjoint two-dimensional Hilbert space $\hat{H}_{pol}=\lbrace\vert V),\vert H)\rbrace$. The full Hilbert space is therefore $\hat{H} = \hat{H}_{path}\otimes \hat{H}_{pol}$.
It can be easily shown \cite{ghose2} that the action of the interferometer shown in Figure 2 is to produce the final state 
\beq
\vert\Phi^+) = \frac{A}{\sqrt{2I_0}}\left[\vert a)\otimes\vert V)+\vert b)\otimes\vert H)\right]\label{phiplus}
\eeq
which is a polarization-path entangled Bell-like state. 

{\flushleft{\em A New Theorem}}

Our aim is to find out whether these classical states are consistent with the notion of noncontextuality, namely the innocuous classical notion that a physical property must be independent of the context in which it is measured. What this implies is that physical systems have properties with predetermined values that are not affected by how the value is measured, i.e. not affected by previous or simultaneous measurement of any other compatible or co-measureable observable. In order to test whether this holds in classical optics, measurements of compatible observables that are not necessarily spatially separated are required. This can be achieved by making a joint measurement of the path and polarization of a single beam. It will now be shown that the correlations between these measurements must satisfy a bound analogous to the CHSH-Bell bound obtained for single-particle mechanics \cite{home} provided the light is prepared in a factorizable state. However, if light is prepared in a nonseparable state like $\vert\Phi^+)$, this bound is violated. 

Noncontextuality of hidden variables used in ontological interpretations of quantum mechanics is a straightforward generalization of the classical notion \cite{ks,mermin, noncontext}. Note, however, that no hidden variables need be invoked in studying classical optical fields because such fields themselves have been amply tested to be deterministic and local as well as noncontextual. The discovery of classical entanglement has changed the situation, and it is important to check whether noncontextuality
holds for such entangled states.

In this context it is necessary to elucidate what one means by a CHSH-like inequality for a classical field theory. Conventional CHSH-Bell inequalities are derived for correlated pairs of {\em particles} that enter distant apparatuses, the particle selecting one of two channels labeled $+1$ and $-1$ in each apparatus. This discreteness is different from quantum discreteness. Quantum mechanics, in fact, is incompatible with Bell's theorem which is a bound that {\em all} theories satisfying local realism must satisfy. 

Particle-like discreteness is altogether absent in classical optics in which intensities of light are measured, and intensities can vary continuously. Nevertheless, since entanglement occurs in classical optics, it is important to check if noncontextuality actually holds generally in classical optics. To do that, it is necessary first to derive a bound for classical optics that follows from noncontextuality. 

We start by defining a correlation
\begin{eqnarray}
E(\theta,\phi)=(\Psi\vert \sigma_{\theta}.\sigma_{\phi}\vert\Psi)
\end{eqnarray}
where $\vert \Psi)$ is an arbitrary normalized classical optical state and
\begin{eqnarray}
\sigma_{\theta}=\sigma_{\theta,0}-\sigma_{\theta,\pi},\\
\sigma_{\phi}=\sigma_{\phi,0}-\sigma_{\phi,\pi},
\end{eqnarray}
with
\begin{eqnarray}
\sigma_{\theta,0}=\frac{1}{2}(|V)+e^{i\theta}|H))((V|+e^{-i\theta}( H|)\otimes I_{path},\nonumber \\
\sigma_{\theta,\pi}=\frac{1}{2}(|V)-e^{i\theta}|H))(( V|-e^{-i\theta}( H|)\otimes I_{path},\nonumber \\
\sigma_{\phi,0}=I_{pol}\otimes\frac{1}{2}(|a)+e^{i\phi}|b))(( a|+e^{-i\phi}( b|),\nonumber \\
\sigma_{\phi,\pi}=I_{pol}\otimes\frac{1}{2}(|a)-e^{i\phi}|b))(( a|-e^{-i\phi}( b|).\label{sigma}
\end{eqnarray}
Hence,
\begin{eqnarray}
\sigma_{\theta}=(e^{-i\theta}|V)( H|+e^{i\theta}|H)( V|)\otimes I_{path},\label{proj1}\\
\sigma_{\phi}=I_{pol}\otimes (e^{-i\phi}|a)( b|+e^{i\phi}|b)( a|).\label{proj2}
\end{eqnarray}
These are projection operators ($\sigma_\theta^2 =1, \sigma_\phi^2 = \mathbb{I}$) that represent polarization and path measurements. Since $\sigma_{\theta}$ and $\sigma_{\phi}$ act upon disjoint Hilbert spaces, they commute.

Let us now consider a general normalized product state 
\begin{eqnarray}
|\Psi)=|\psi_{pol})|\psi_{path})=(\cos\alpha|V)+e^{i\beta}\sin\alpha|H))(\cos\gamma|a)+e^{i\delta}\sin\gamma|b))
\end{eqnarray}
where $\alpha, \beta, \gamma, \delta$ are arbitrary parameters.
The correlation for such a state is
\begin{eqnarray}
E(\theta,\phi)&=&(\psi_{pol}|\sigma_{\theta}|\psi_{pol})(\psi_{path}|\sigma_{\phi}|\psi_{path})\nonumber \\
&=& E_{pol}(\theta)E_{path}(\phi),\label{exp}
\end{eqnarray}
with
\begin{eqnarray}
E_{pol}(\theta)=\sin\alpha\cos(\beta-\theta),\label{epol}\\
E_{path}(\phi)=\sin\gamma\cos(\delta-\phi).\label{epath}
\end{eqnarray}
\beq
-1\leqslant E(\theta,\phi)\leqslant 1.
\eeq
Let us define a quantity $S$ by
\ben
S(\theta_{1},\phi_{1};\theta_{2},\phi_{2}) &=& E(\theta_{1},\phi_{1})+E(\theta_{1},\phi_{2})-E(\theta_{2},\phi_{1})+E(\theta_{2},\phi_{2})\nonumber\\
&\equiv& m(b+a)+n(b-a)
\een
where
\begin{eqnarray}
m=E_{pol}(\theta_{1}),\,\,  n=E_{pol}(\theta_{2}),\nonumber \\
a=E_{path}(\phi_{1}),\,\,   b=E_{path}(\phi_{2}).
\end{eqnarray}
If one makes use of the triangle inequality 
\begin{eqnarray}
|S|\leq |m||b+a|+|n||b-a|
\end{eqnarray}
and the bounds $|m|,|n|\leq 1$, one can easily see that
\begin{eqnarray}
|S|\leq |m||b+a|+|n||b-a|\leq |b+a|+|b-a|.
\end{eqnarray}
Using
\begin{eqnarray}
{\rm max}(A,B)=\frac{A+B}{2}+\frac{|A-B|}{2},
\end{eqnarray}
one can easily verify that
\beq
|S|\leq 2\label{bound}
\eeq
for all possible combinations of values of $a,\,b \leq \pm 1$.
This result depends only on the fact that {\em the correlations lie between $-1$ and $+1$} (guaranteed by the results (\ref{epol}) and (\ref{epath})) and {\em no discreteness assumption is necessary}. Hence, this is a new and non-trivial result for classical optics, analogous to CHSH-Bell inequalities in particle mechanics.

Let us now see calculate the correlation for the normalized state (\ref{phiplus}),
\ben
E(\theta,\phi)&=&(\Phi^+|\sigma_{\theta}\cdot\sigma_{\phi}|\Phi^+)\nonumber\\
&&\:=( \Phi^+\vert \,[(+)\sigma_{\theta,0} + (-)\sigma_{\theta,\pi} ]. [(+)\sigma_{\phi,0} + (-)\sigma_{\phi,\pi} ]\vert\Phi^+).
\een
To measure this correlation, one must measure the intensities of light at the final detector in path $a_{out}$ (Fig. 2) corresponding to four possible combinations of settings of the phase-shifter $PS_b$ and the polarization rotator $PR_b$ in path b as follows:
\ben
E(\theta, \phi) &=& (\Phi^+|[\sigma_{\theta,0}\cdot\sigma_{\phi,0}+\sigma_{\theta,\pi}\cdot\sigma_{\phi,\pi}\nonumber\\ &&\: -\sigma_{\theta,0}\cdot\sigma_{\phi,\pi}-\sigma_{\theta,\pi}\cdot\sigma_{\phi,0}]|\Phi^+),\label{E} 
\een
with the intensities given by
\ben
I(\theta,\phi)&=&( \Phi^+|\sigma_{\theta,0}\cdot\sigma_{\phi,0}|\Phi^+),\nonumber\\
I(\theta+\pi,\phi+\pi)&=&( \Phi^+|\sigma_{\theta,\pi}\cdot\sigma_{\phi,\pi}|\Phi^+),\nonumber\\
I(\theta+\pi,\phi)&=&( \Phi^+|\sigma_{\theta,\pi}\cdot\sigma_{\phi,0}|\Phi^+),\nonumber\\
I(\theta,\phi+\pi)&=&( \Phi^+|\sigma_{\theta,0}\cdot\sigma_{\phi,\pi}|\Phi^+).\nonumber\\
\een
It is clear that $I(\theta,\phi)=\frac{1}{2}[1+\cos(\theta+\phi)]$ from (\ref{phiplus}) and the definitions (\ref{sigma}), as one would expect from a classical optical state. One can rewrite $E(\theta, \phi)$ in terms of the normalized intensities as
\ben
E(\theta,\phi)&=&\dfrac{I(\theta,\phi)+I(\theta+\pi,\phi+\pi)-I(\theta+\pi,\phi)-I(\theta,\phi+\pi)}{I(\theta,\phi)+I(\theta+\pi,\phi+\pi)+I(\theta+\pi,\phi)+I(\theta,\phi+\pi)}\nonumber\\
 &=& \cos(\theta+\phi).
\een
Notice that this is not in a product or factorizable form. It follows from this that the noncontextuality bound (\ref{bound}) is violated by the state $\vert \Phi+)$ for the set $\theta_1 = 0, \theta_2 = \pi/2, \phi_1 = \pi/4, \phi_2 = - \pi/4$ for which $S = 2\sqrt{2}$. There is no nonlocality in this result because the path and polarization changes are made on the same state in path $b$. This result shows that the path and polarization of classical light in entangled states like $\vert \Phi^+)$ are {\em contextual}, i.e. the measurement of path/polarization depends on the measurement of polarization/path although they are compatible properties. This is analogous to the test of noncontextuality carried out using neutron interferometry \cite{hasegawa}.
\vskip 0.1in

{\flushleft{\em Violation of the axioms of the Kochen-Specker Theorem}}

\vskip 0.1in
It is possible to show that classical entangled states like the ones considered above can violate the axioms of the Kochen-Specker theorem. Let us follow the simple approach of Mermin and Peres \cite{mermin2} and consider the state
\beq
\vert\Phi^-) = \frac{A}{\sqrt{2I_0}}\left[\vert a)\otimes\vert V) - \vert b)\otimes\vert H)\right].
\eeq
Consider also the six hermitian operators $J_x^{pol}, J_x^p, J_y^{pol}, J_y^p, J_x^{pol} J_y^p, J_y^{pol} J_x^p$  which can act on this state, where $J^{pol}$s are Jones matrices acting on $\hat{H}_{pol}$ and $J^p$s are analogous matrices acting on $\hat{H}_{path}$. Clearly, $J^{pol}$s and $J^p$s commute. Notice that all the operators mutually commute except the last two. These matrices have eigenvalues $\pm 1$.
Since 
\ben
J_x^p\vert a) &=& \vert b),\,\,\,\,\,\, J_x^p\vert b) = \vert a),\\
J_y^p \vert a) &=& i\vert b),\,\,\,\,\,\, J_y^p \vert b) = -i\vert a),\\
J_x^{pol}\vert V) &=& \vert H),\,\,\,\,\,\, J_x^{pol}\vert H) = \vert V),\\
J_y^{pol}\vert V) &=& -i\vert H),\,\,\,\,\,\, J_y^{pol}\vert H) = i\vert V),
\een
it follows that
\ben
J_x^{pol}\,.\, J_x^p \vert\Phi^-) &=& -\vert\Phi^-),\\
J_y^{pol}\,.\, J_y^p \vert\Phi^-) &=& -\vert\Phi^-),\\
J_x^{pol} J_y^p\, .\, J_x^{pol}\,.\,J_y^p \vert\Phi^-) &=& +\vert\Phi^-),\\
J_y^{pol} J_x^p\, .\, J_y^{pol}\,.\,J_x^p \vert\Phi^-) &=& +\vert\Phi^-),\\
J_x^{pol} J_y^p\, .\, J_y^{pol}J_x^p \vert\Phi^-) &=& -\vert\Phi^-).
\een
This shows that the state $\vert\Phi^-)$ is an eigenstate of the operators on the left-hand sides with eigenvalues $\pm 1$.
Since each of the six operators occurs exactly twice on the left-hand sides, the product of the left-hand sides is $+1$ if it is assumed that each property of the state has a predetermined and context independent value $\pm 1$. However, notice that the product of the right-hand sides is $-1$. This is an obvious logical contradiction. Since the Kochen-Specker theorem is based on the two axioms of (i) value definiteness and (ii) noncontextuality, these assumptions cannot hold for entangled classical states like $\vert\Phi^-)$. 

\section{Nonseparable Two-mode Classical States and Local Realism}

Consider a laser beam in the double mode $\vert H)^a \otimes \vert H)^b$ where each bracket represents a first-order Hermite-Gaussian transverse mode $\psi_H(\vec{r})^{a/b} \hat{e}_H$ with horizontal orientation in a path ($a$ or $b$), $\hat{e}_H$ being the horizontal ($H$) linear polarization unit vector. Similarly, let $\vert V)^c \otimes \vert V)^d$ be a second laser double mode with vertical orientation in paths $c$ and $d$. They are product states. Particular cases for $\psi_H(\vec{r})$ and $\psi_V(\vec{r})$ would be the modes ${\rm TEM}_{01(y)}$ and ${\rm TEM}_{01(x)}$ respectively. It is possible to combine these beams and produce nonseparable linear combinations as shown in Figure 3. Let us assume that the four modes have a common intensity $I_0 = \vert A\vert^2$. Consider one such state
\beq
\vert \Phi^+) = \frac{A}{\sqrt{2I_0}} \left[\vert H)^a \otimes \vert H)^b + \vert V)^a \otimes \vert V)^b \right].\label{bell} 
\eeq
It is an analog of the two-particle Bell state $\vert \Phi^+\rangle$ in quantum mechanics. Clearly, all four Bell-like states can be produced in this way. Let us see if they satisfy local realism. 
\begin{figure}[h!]
\centering
{\includegraphics[scale=5]{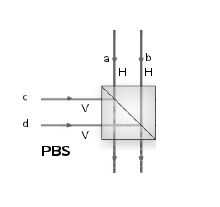}}
\caption{\label{Figure 3}{\footnotesize Combining two double-mode beams, one in the VV and the other in the HH mode.}}
\end{figure}

Let a polarization analyzer be placed in each path such that their polarization axes make arbitrary angles $\alpha$ and $\beta$ with the $x$ axis, and let $\alpha - \beta = \theta$. Let us define the correlation
\begin{eqnarray}
E(\alpha,\beta)=(\Psi\vert \sigma^{a}_{\alpha}\otimes\sigma^{b}_{\beta}\vert\Psi)\\
\end{eqnarray}
with 
\begin{eqnarray}
\sigma^{a}_{\alpha}=2P^{a}_{\alpha}-I,\\
\sigma^{b}_{\beta}=2P^{b}_{\beta}-I,
\end{eqnarray}
and
\begin{eqnarray}
P^{a}_{\alpha}=[\cos\alpha |H^{a})+\sin\alpha |V^{a})][\cos\alpha (H^{a}|+\sin\alpha (V^{a}|],\nonumber \\
P^{a}_{\beta}=[\cos\beta |H^{b})+\sin\beta |V^{b})][\cos\beta (H^{b}|+\sin\beta (V^{b}|],
\end{eqnarray}
the projection operators representing the two polarization analyzers.
Hence,
\begin{eqnarray}
\sigma^{a}_{\alpha}=\cos 2\alpha[|H^{a})(H^{a}|-|V^{a})(V^{a}|] +\sin 2\alpha[|H^{a})(V^{a}|+|V^{a})(H^{a}|],\\
\sigma^{b}_{\beta}=\cos 2\beta[|H^{b})(H^{b}|-|V^{b})(V^{b}|] +\sin 2\beta[|H^{b})(V^{b}|+|V^{b})(H^{b}|].
\end{eqnarray}
Writing an arbitrary normalized product state as 
\begin{eqnarray}
|\Psi)=|\psi^{a})|\psi^{b})=[\cos\theta|H^{a})+e^{i\phi}\sin\theta|V^{a})][\cos\rho|H^{b})+e^{i\eta}\sin\rho|V^{b})],
\end{eqnarray}
one finds
\begin{eqnarray}
E(\alpha,\beta)&=&(\psi^{a}|\sigma^{a}_{\alpha}|\psi^{a})(\psi^{b}|\sigma^{b}_{\beta}|\psi^{b})\nonumber \\
&=& E^{a}(\alpha)E^{b}(\beta).
\end{eqnarray}
This shows that the expectation values 
\begin{eqnarray}
E^{a}(\alpha)=\cos 2(\alpha+\theta)\sin^{2}\frac{\phi}{2}+\cos 2(\alpha-\theta)\cos^{2}\frac{\phi}{2},\\
E^{b}(\beta)=\cos 2(\alpha+\rho)\sin^{2}\frac{\eta}{2}+\cos 2(\alpha-\rho)\cos^{2}\frac{\eta}{2},
\end{eqnarray}
measured by the polarization analyzers in paths (a) and (b) are independent, and that local realism holds for such states. Note that the conditions
\ben
-1\leqslant E^{a}(\alpha)&\leqslant 1,\nonumber\\ -1\leqslant E^{b}(\beta)&\leqslant 1, \label{b}
\een
must hold. Hence, if one defines the quantity
\begin{eqnarray}
S^\prime(\alpha_{1},\beta_{1};\alpha_{2},\beta_{2})=E(\alpha_{1},\beta_{1})+E(\alpha_{1},\beta_{2})-E(\alpha_{2},\beta_{1})+E(\alpha_{2},\beta_{2}),
\end{eqnarray}
then, using the arguments in the previous section, it is at once clear that that the bound
\beq
\vert S^\prime \vert \leq 2\label{sbound} 
\eeq
must hold for product states. It is also clear that no discreteness assumption is required to derive this fundamental result.

Consider now the entangled state $|\Phi^{+})$ (\ref{bell}) for which the correlation function is
\begin{eqnarray}
E(\alpha,\beta)&=&(\Phi^{+}|\sigma^{a}_{\alpha}\otimes\sigma^{b}_{\beta}|\Phi^{+}).\nonumber \\
			   &=&  \cos2\theta,
\end{eqnarray} 
which is not in the form of a product of expectation values. It is easy to see that the bound (\ref{sbound}) is violated b this states for the set
$\alpha_{1}=0$ , $\alpha_{2}=\dfrac{\pi}{4}$, $\beta_{1}=\dfrac{\pi}{8}$, $\beta_{2}=-\dfrac{\pi}{8}$ for which $\vert S^\prime \vert = 2\sqrt{2}$.

Does that mean that such states violate locality? The answer is decidedly no, because there is no collapse of the state on local measurements of polarization, as in standard quantum mechanics, and hence no nonlocal effect. All that happens in a classical measurement is that one of the polarization states is projected (i.e. selected) and recorded while the other is simply blocked--it does not disappear as in projective quantum measurement. Violation of the inequality therefore only shows that, like their counterparts in quantum mechanics, these states are more correlated than expected from traditional classical notions of local realism.

Multimode lasers can be used in principle to produce multimode nonseparable states of classical light.

\section{Influence of the Pancharatnam Phase on a Classical Bell-like State}

When the polarization of a light beam is rotated in a cyclic path, the polarization vector comes back with an overall phase that is the sum of the dynamical phase and the Pancharatnam phase \cite{pancha} which is half the solid angle subtended by the circuit at the centre of the Poincare sphere. To isolate the Pancharatnam phase, the effect of the dynamical phase has to be isolated, and this can be done in the following manner. Let a half-wave plate ($HWP$) be inserted between two co-aligned quarter-wave plates ($QWP$s) to form a combination $QHQ$. Let such a device be inserted into one arm of an interferometer. It can be shown \cite{love} that if plane polarized light of phase $\alpha$ is incident on the first $QWP$ at 45$^0$ to its axes, its phase shifts to $\alpha + \pi - 2\gamma$ where $\pi$ is the dynamical phase shift due to the optical path lengths and $\gamma$ is the angle between the axes of the $HWP$ and the $QHQ$. As the $HWP$ is rotated, $\gamma$ changes, and this change is detectable through fringe shifts of $-2\gamma$. This is a pure geometrical phase shift brought about by polarization changes, and is hence the Pancharatnam phase. The closed circuit on the Poincare sphere is shown in Figure 4.
\begin{figure}[ht]
\centering
{\includegraphics[scale=0.3]{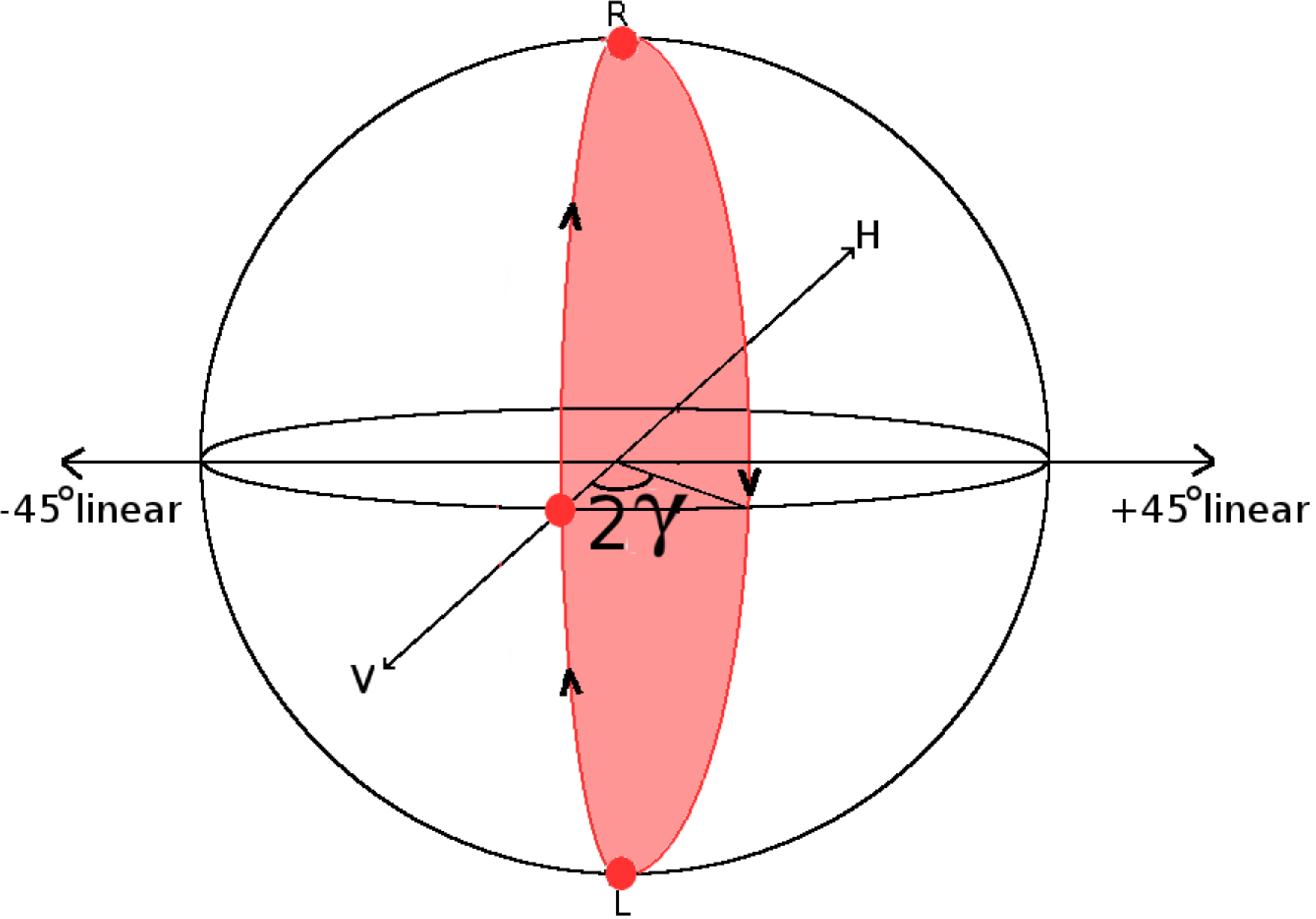}}
\caption{\label{Figure 4}{\footnotesize The closed circuit on the Poincare sphere}}
\end{figure}

Let us now consider the experimental arrangement shown in Figure 5 which is a modification of Fig. 2. Let the first beam splitter $NPBS_1$ be replaced by a variable asymmetric non-polarizing beam splitter $VNPBS_1$ with the reflection and transmission coefficients $r = \sin \theta_1$ and $t = \cos \theta_1$, and let a $QHQ$ and a half-wave plate $HWP_a$ be inserted in path $a$ and a full-wave plate $FWP_b$ and a half-wave plate $HWP_b$ be inserted in path $b$. Fixing the initial arrangement so that $\theta_1 = \pi/4$, the state after $QHQ$ in path $a$ and $FWP_b$ in path $b$ s
\beq
\vert \Phi_P\rangle = \frac{A e^{i\pi}}{\sqrt{2I_0}}[e^{-2i\gamma}\vert a\rangle\vert V\rangle + \vert b\rangle\vert H\rangle].
\eeq
Then, after passage through the other optical elements but before the second beam splitter $NPBS_2$ and for variable $\theta_1$, the state is given by
\begin{figure}
{\includegraphics[scale=0.6]{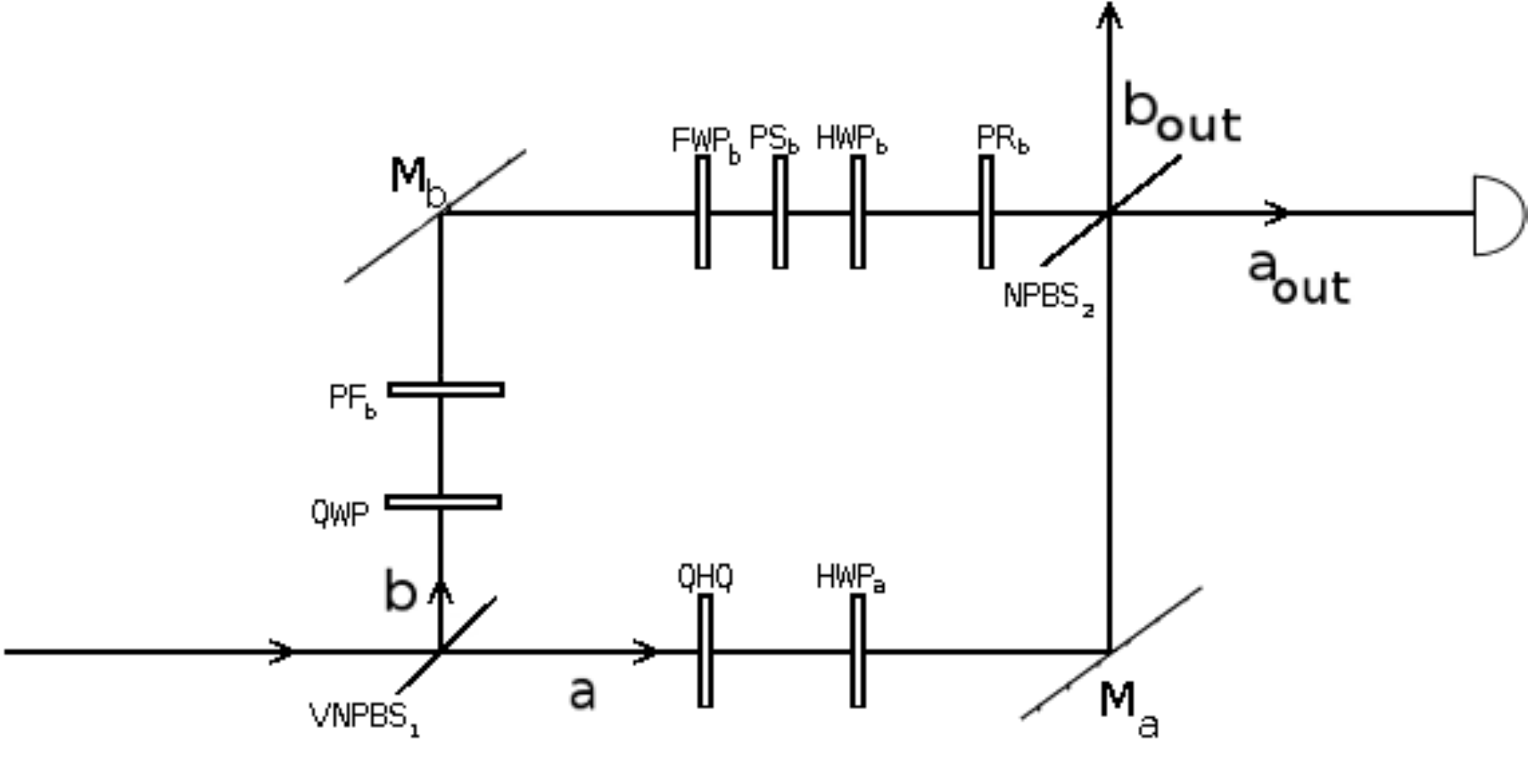}}
\caption{\label{Figure 5} {\footnotesize Fig. 2 modified by replacing $NPBS_1$ by a variable $VNPBS_1$ and adding optical elements $QHQ$ and $HWP_a$ in path $a$ and a $FWP_b$ and $HWP_b$ in path $b$.}}
\end{figure}

\ben
\vert\Phi\rangle &=& \frac{Ae^{i\pi}}{\sqrt{I_0}}[e^{-i2\gamma}\cos\theta_1\vert a\rangle\otimes\left(\sin\theta_2\vert H\rangle -\cos\theta_2 \vert V\rangle\right)\nonumber\\ &+& e^{i\phi}\sin\theta_1\vert b\rangle\otimes\left(\cos\theta_2\vert H\rangle + e^{i\theta}\sin\theta_2 \vert V\rangle\right)]
\een
where $HWP_a$ and $HWP_b$ have their fast axes at $\theta_2/2$ to the vertical.
After passing $NPBS_{2}$, the final state is
\ben
\vert\Phi\rangle &=& \frac{Ae^{i\pi}}{\sqrt{I_0}}[\vert a\rangle\otimes \left\{\cos\theta_1\sin\theta_2 +ie^{i\phi}\sin\theta_1\cos\theta_2\right)\vert H\rangle\nonumber\\ &+& \left(ie^{i(\theta + \phi)}\sin\theta_2\sin\theta_1- e^{-2i\gamma}\cos\theta_1\cos\theta_2\right\}\vert V\rangle\nonumber\\ &+& \vert b\rangle\otimes\left\{\left(i e^{-2i\gamma}\cos\theta_1\sin\theta_2 + e^{i\phi}\sin\theta_1\cos\theta_2\right)\vert H\rangle\right)\nonumber\\ &+& \left(e^{i(\theta + \phi)}\sin\theta_2\sin\theta_1 - i e^{-2i\gamma}\cos\theta_1\cos\theta_2\right)\vert V\rangle\}].
\een
Therefore, using the projection operators (\ref{proj1}) and (\ref{proj2}) 
with
\ben
\vert {\rm path}\rangle &=&\cos\theta_{1}\vert a\rangle + e^{i\phi}\sin\theta_{1}\vert b\rangle,\\
\vert {\rm pol}\rangle &=& \cos\theta_{2}\vert V\rangle + e^{i\theta}\sin\theta_{2}\vert H\rangle,
\een
we get
\beq
E\left(\theta_{1},\theta_2,\theta,\phi, \gamma \right)=\langle \sigma_\theta\,.\,\sigma_\phi\rangle = \cos 2\theta_{1}\cos 2\theta_{2} + \cos (\theta + \phi -2\gamma) \sin 2\theta_{1}\sin 2\theta_{2}.
\eeq
One can then define the function 
\ben
S &=&  E (\theta_{1},\theta_{2},\theta,\phi, \gamma) - E (\theta_{1}^{'},\theta_{2},\theta,\phi^{'},\gamma) + E (\theta_{1},\theta_{2}^{'},\theta^{'},\phi,\gamma)\nonumber\\ &+& E (\theta_{1}^{'},\theta_{2}^{'},\theta^{'},\phi^{'},\gamma)
\een
which must satisfy the Bell-like inequality $-2 \leq S \leq 2$.
Putting $\theta_{2}=0$ and $\theta=0$ we have
\ben
S&=& \cos\left(2\theta_{1}\right) - \cos2\theta'_{1} + \cos\left(2\theta_{1}\right)\cos\left(2\theta'_{2}\right)+
\cos\left(\theta'+\phi-2\gamma\right)\sin\left(2\theta_{1}\right)\sin\left(2\theta'_{2}\right)\nonumber\\ 
&+&\cos 2\theta'_{1}\cos\left(2\theta'_{2}\right)+
\cos\left(\theta'+\phi'-2\gamma\right)\sin\left(2\theta'_{1}\right)\sin\left(2\theta'_{2}\right).
\een
Setting the polar bell-angles at $\theta'_{1}=\dfrac{3\pi}{8},\theta_{1}=\dfrac{\pi}{8},\theta'_{2}=\dfrac{\pi}{4}$, one gets 
\beq
S=\sqrt{2}+\frac{1}{\sqrt{2}}\left[\cos\left(\theta'+\phi'-2\gamma\right)+\cos\left(\theta'+\phi-2\gamma\right)\right].
\eeq
With $\phi'=\phi$ and $\theta'+\phi=2\gamma$, $S=2\sqrt{2}$, i.e. there is maximal violation of the Bell-like inequality. 

\begin{figure}{ht}
{\includegraphics[scale=0.4]{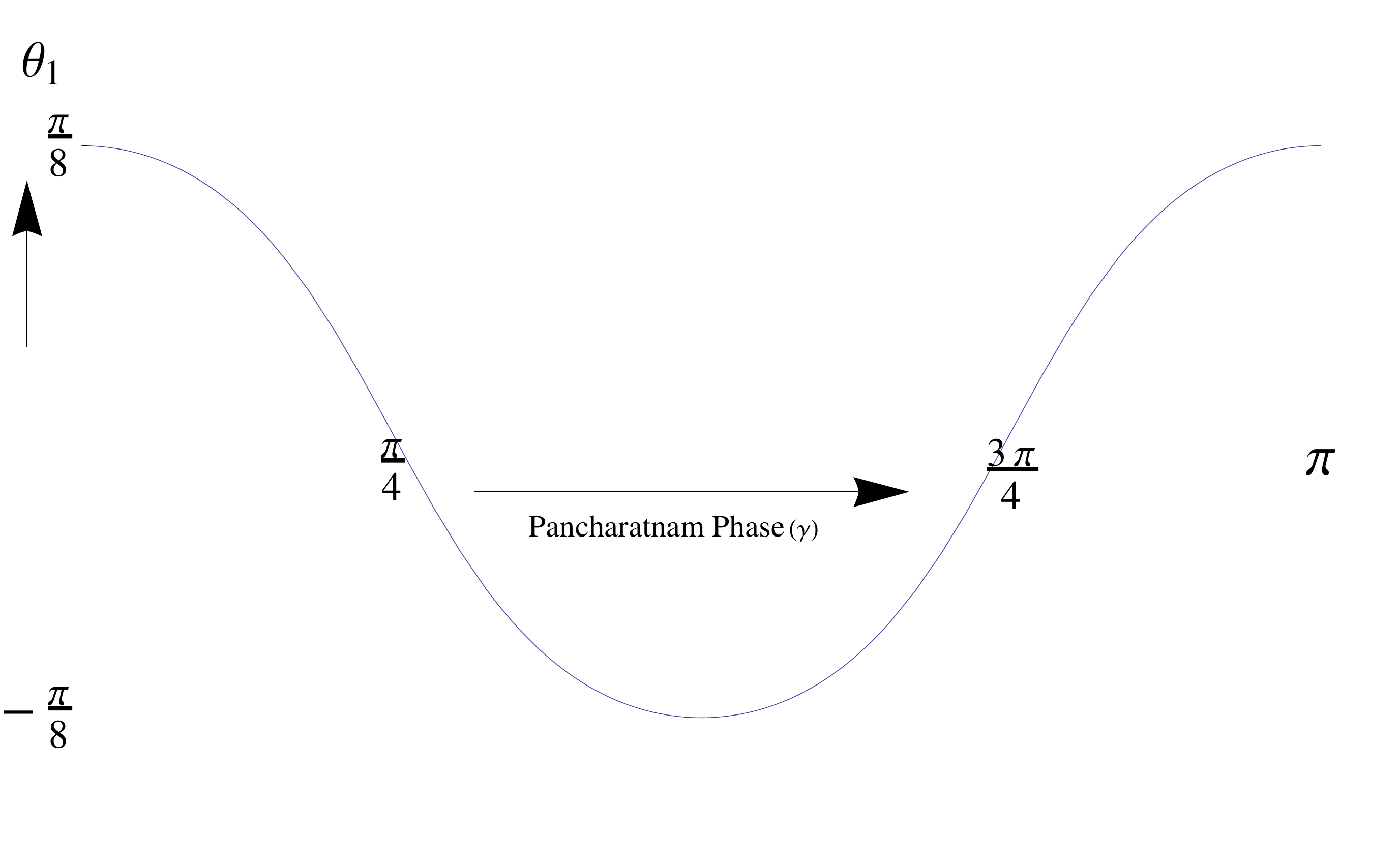}}
\caption{\label{Figure 6}{\footnotesize The polar angle $\theta_1$ is plotted against the Pancharatnam phase $\gamma$}}
\end{figure}

\begin{figure}{ht}
{\includegraphics[scale=0.4]{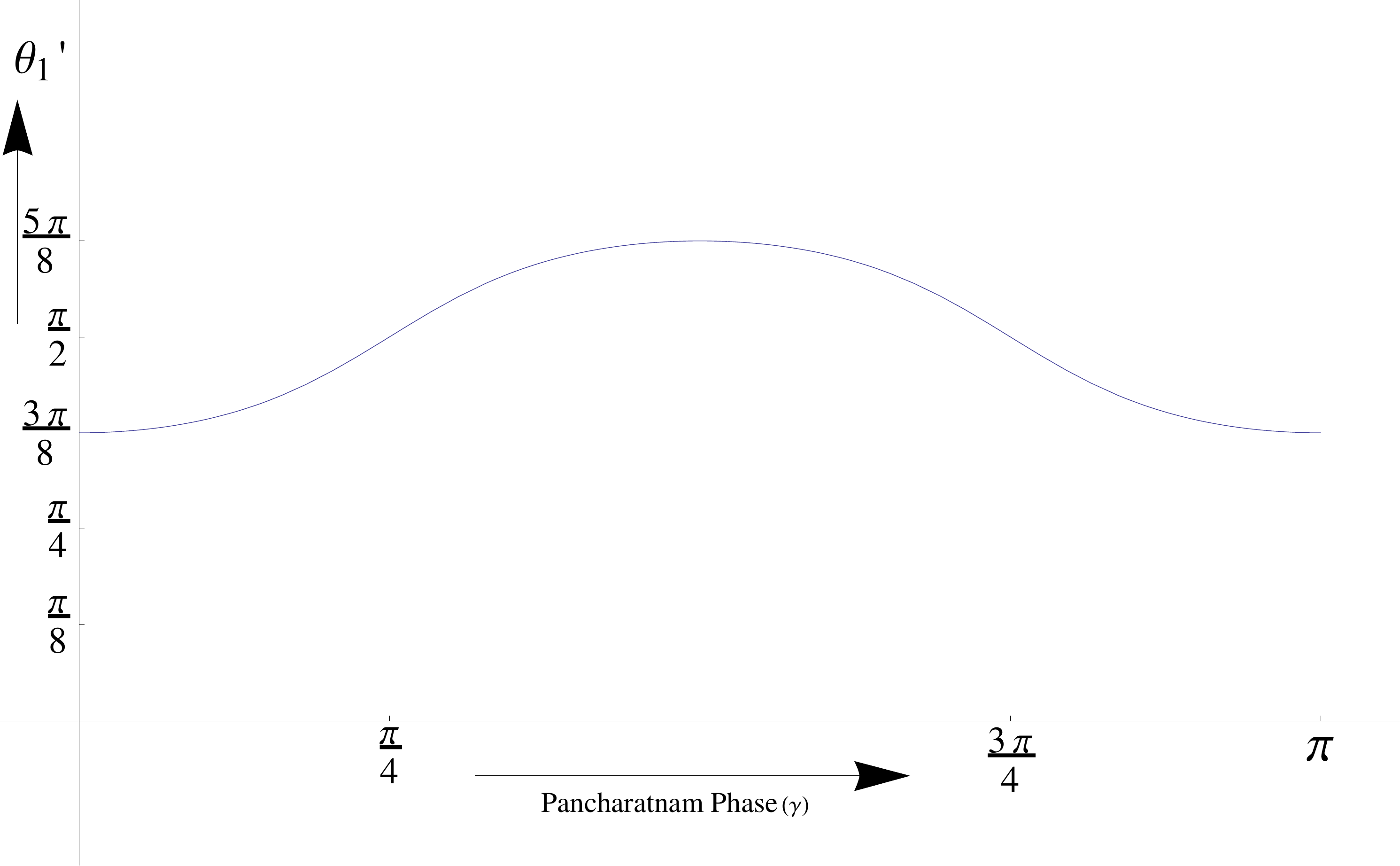}}
\caption{\label{Figure 7}{\footnotesize The polar angle $\theta_1^{'}$ is plotted against the Pancharatnam phase $\gamma$}}
\end{figure}

\begin{figure}{ht}
{\includegraphics[scale=0.6]{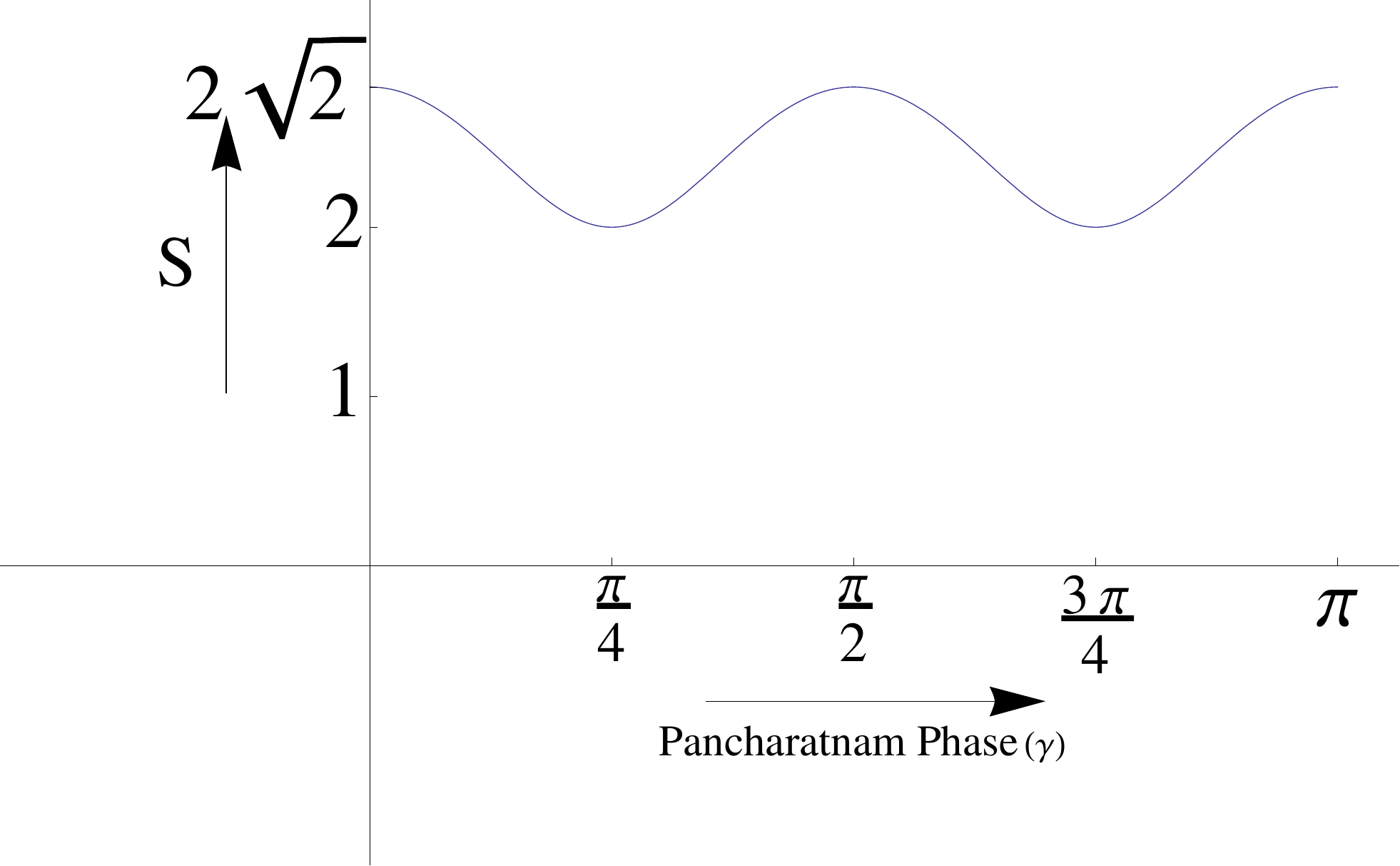}}
\caption{\label{Figure 8}{\footnotesize The extrema of the S function as a function of the Pancharatnam phase for fixed azimuthal angles $\left\lbrace\theta'=\theta=\phi'=\phi=0\right\rbrace$}}
\end{figure}

Let us next look at the reverse problem, i.e. calculate the extremum values of $S$ as a function of the Pancharatnam phase $\gamma$ for fixed azimuthal angles $\left\lbrace\theta'=\theta=\phi'=\phi=0\right\rbrace$. Then,
\ben
S&=& \cos 2\theta_1 - \cos 2\theta'_1 + \cos 2\theta'_2\left(\cos 2\theta_1 + \cos 2\theta'_1\right)\nonumber\\  &+&
\cos 2\gamma\sin 2\theta'_2\left(\sin 2\theta_1 +
\sin 2\theta'_1\right).
\een
By requiring the partial derivatives of $S$ w.r.t $\theta_{1},\theta'_{1},\theta'_{2},$ to vanish,
\ben
\frac{\partial S}{\partial \theta_{1}}&=& -2\sin 2\theta_1 -2\sin 2\theta_1\cos 2\theta'_2 + 2\cos 2\gamma\sin 2\theta'_2\cos 2\theta_1=0,\\
\frac{\partial S}{\partial \theta'_{1}}&=& - 2\sin 2\theta'_1\cos 2\theta'_2 + 2\sin 2\theta_1' + 2\cos 2\theta_1\cos 2\gamma\sin 2\theta'_2=0,\\
\frac{\partial S}{\partial \theta'_{2}}&=& -2\sin 2\theta'_2(\cos 2\theta_1 + \cos 2\theta'_1)\nonumber\\ &+&  2\cos 2\gamma\cos 2\theta'_2(\sin 2\theta_1 + \sin 2\theta'_1)=0, 
\een
one gets 
\ben
\theta_{1}&=&\frac{1}{2}\arctan\left(\cos\left(2\gamma\right)\right),\\
\theta_{1}^{'}&=&\frac{\pi}{2}-\theta_{1},\\
\theta'_{2}&=&\frac{\pi}{4}
\een
as the angles for which $S$ is extremal. The results are shown in Figures 6-8. The function $S$ exceeds the Bell limit +2 for all values of $\gamma$ except at the points where it has its minima (Fig. 8). 

\section{Information Processing with Classical Light}

We will now discuss to what extent classical polarization optics can be used to simulate quantum information processing tasks \cite{sen}, \cite{spreeuw}. This is very important from a practical point of view because coherence and entanglement are robust in classical optics, thus circumventing the problem of decoherence in quantum information processing. 
However, as already emphasized, there are certain fundamental differences between classical optical coherence and entanglement and their quantum counterparts, namely (a) there is no nonlocality in classical optics, and (b) there is no `no-cloning theorem' \cite{nocloning} in classical optics. As a result, error correction protocols have to be specially designed. The exponential resources problem with classical optics also needs further investigation.

We will now show how the simple quantum logic gates and networks \cite{bell} can be realized with classical optics.

\subsection*{Gates analogous to single qubit gates}

Let us first consider the case of a linearly polarized classical light beam passing through a half-wave plate HWP with its fast axis inclined at($\frac{\phi-\theta}{2}$) to the vertical (Fig. 9). The transition from the initial state $\theta$ to the final state $\phi$ of polarization is given by 
\begin{eqnarray}
J\left((\phi-\theta)/2\right)|\theta\rangle&=&|\phi\rangle, \nonumber\\
J((\phi-\theta)/2) &=&\left(\begin{array}{cc}
 										\cos(\phi-\theta)&	 \sin(\phi-\theta)\\
 										\sin(\phi-\theta)&	-\cos(\phi-\theta)		
 										\end{array}
 									\right)\nonumber\\
\vert\theta\rangle&=&\left(\begin{array}{c}
 										\cos\theta\\
 										\sin\theta
 									  \end{array}
 								\right)	\nonumber  											
\end{eqnarray}
where $J$ is a Jones matrix. More generally, the Jones vector $\vert \theta\rangle$ can also be written in the form
\beq
\vert \theta\rangle = c_0 \vert 0\rangle + c_1 \vert 1\rangle \label{cebit}
\eeq
with $c_0$ and $c_1$ classical complex amplitudes and $\vert 0\rangle$ and $\vert 1\rangle$ denoting the basis states
\beq
\vert 0\rangle = \left(\begin{array}{c}
 										1	\\
 										0		
 										\end{array}
 									\right),\,\,\,\,\,\,\,\,\,\,\vert 1\rangle =\left(\begin{array}{c}
 										0	\\
 										1		
 										\end{array}
 									\right),
\eeq 
so that
\beq
\vert \theta\rangle = \left(\begin{array}{c}
 										c_0	\\
 										c_1
 										\end{array}\right).
\eeq

\begin{figure}[ht]
\centering
{\includegraphics[scale=0.5]{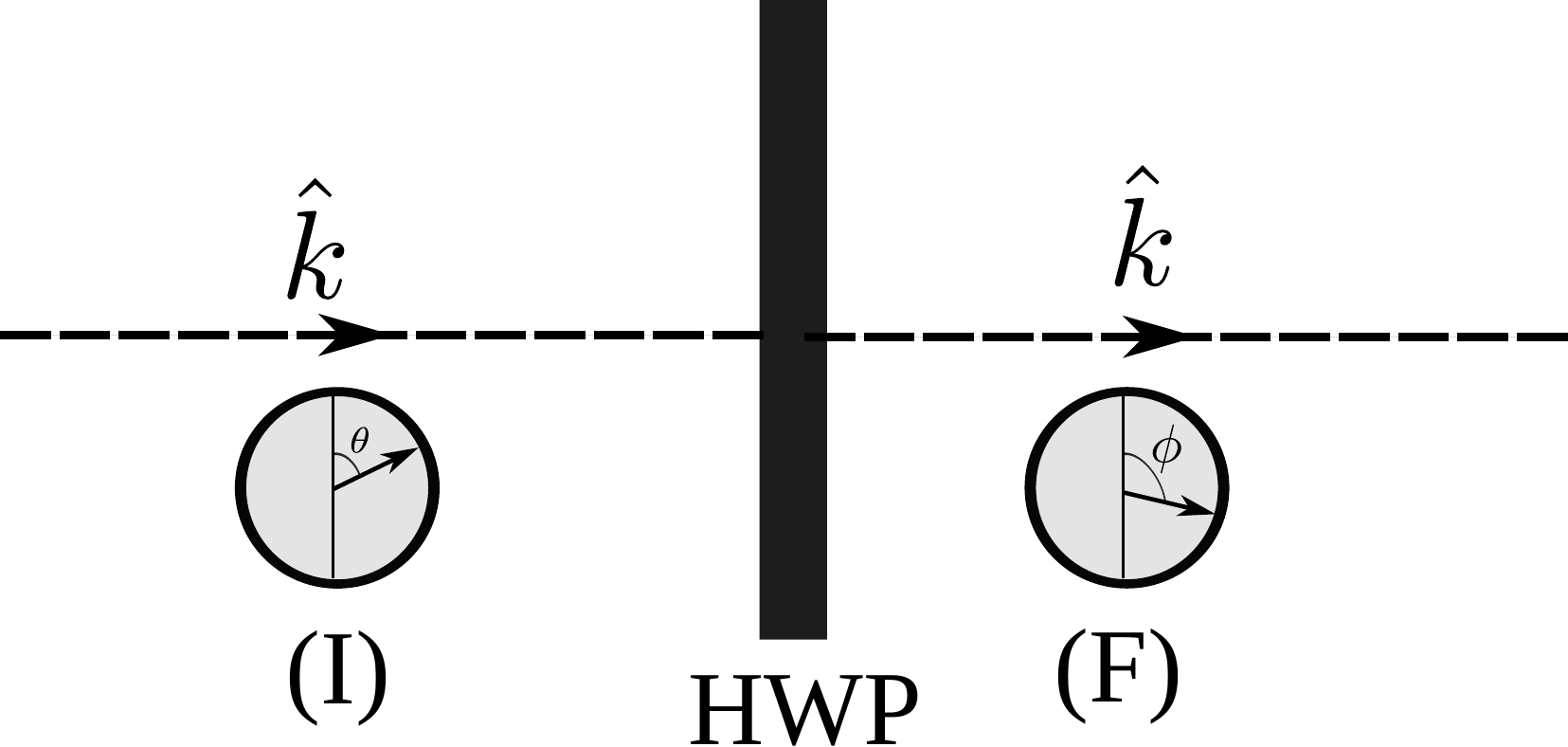}}
\caption{\label{Figure 9}{\footnotesize The action of a half-wave plate HWP with its fast axis inclined at($\frac{\phi-\theta}{2}$) to the vertical on a linearly polarized light beam. (I) and (F) represent the initial and final states of polarization respectively.}}
\end{figure}

The coefficients $\vert c_0\vert^2$ and $\vert c_1\vert^2$ are the intensities measured by photodetectors. Then (\ref{cebit}) is the classical analog of a qubit, and may be called a polarization {\em cebit}. A similar relation holds for path cebits with $\vert 0\rangle$ and $\vert 1\rangle$ denoting two orthogonal paths. In the classical optical case, though the path and polarization cebits are locally specified, they are independent degrees of freedom, being elements of disjoint Hilbert spaces. They can therefore be used as control or target cebits. 

\subsection*{Gates analogous to two qubit unitary gates}

\subsubsection*{The CNOT Gate}
Let us first consider the propagation mode or the path as the control cebit and the polarization as the target cebit. After the beam splitter NPBS (Fig. 10) the state of an incident H beam is

\begin{figure}[ht]
\centering
\hspace*{1.6cm}
\includegraphics[scale=0.7]{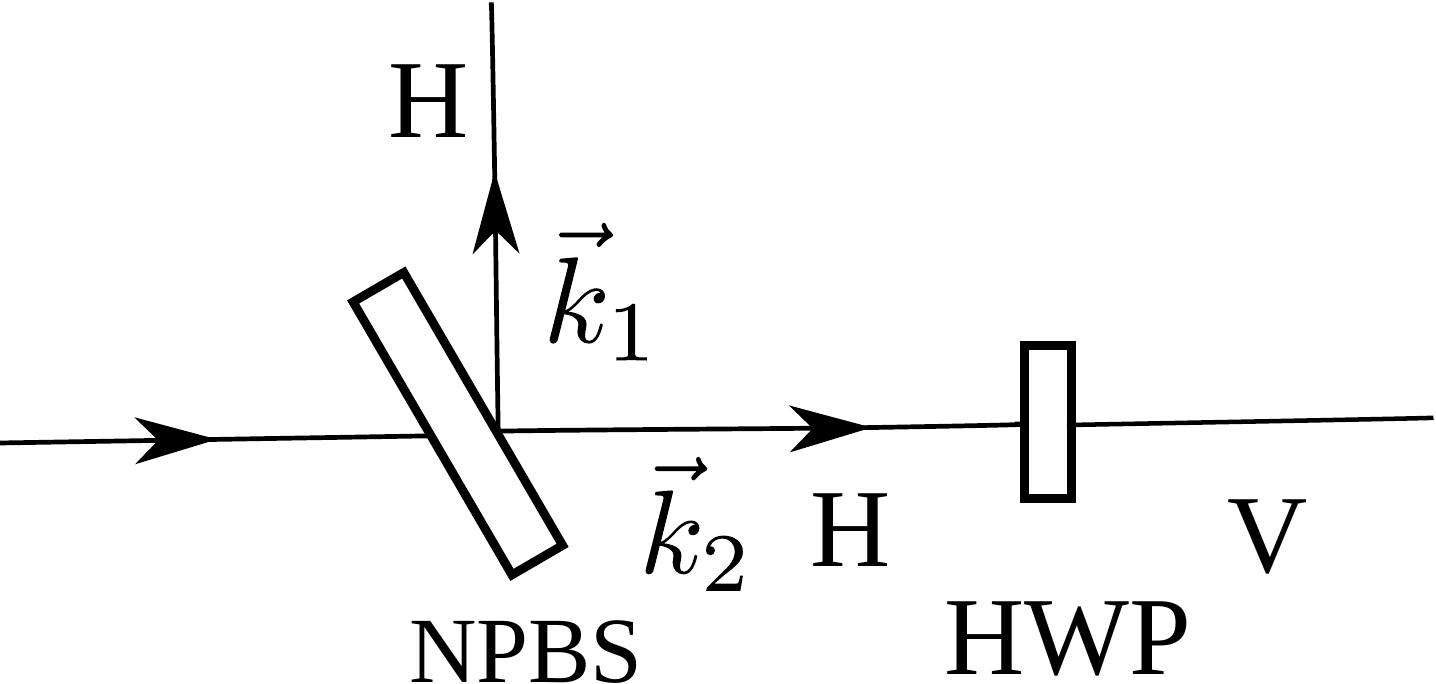} 
\caption{\label{Figure 10}{\footnotesize A CNOT gate with path as the control cebit. The fast axis of the half-wave plate HWP is oriented at $45^{0}$ to the vertical.}}
\end{figure}

\begin{equation}
|\Psi\rangle=(|\vec{k}_{1}\rangle +|\vec{k}_{2}\rangle)|H\rangle
\end{equation}
If a half-wave plate HWP is placed in the path $\vec{k}_{2}$, it flips the H polarization state to the V polarization state. Therefore the final state is
\begin{equation}
|\Psi\rangle=|\vec{k}_{1}\rangle |H\rangle +|\vec{k}_{2}\rangle |V\rangle
\end{equation}
This is a CNOT gate as it flips the polarization of the target cebit only if the control cebit is the path $\vec{k}_{2}$ and not $\vec{k}_{1}$. Since the intensity remains the same, we have a succesful realization of CNOT. 

\begin{figure}[ht]
\centering
\hspace*{1.6cm}
\includegraphics[scale=1.0]{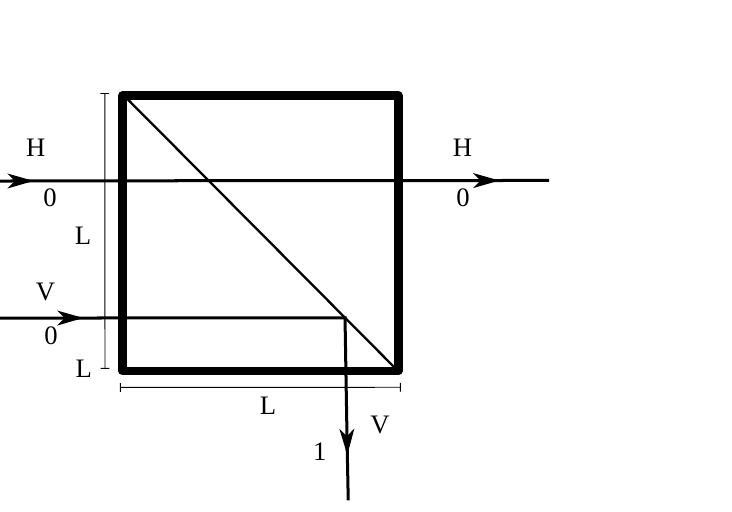} 
\caption{\label{Figure 11}{\footnotesize A CNOT gate with polarization as the control bit. PBS is a polarizing beam splitter .}}
\end{figure}

One can also have a CNOT gate with polarization as the control cebit to fix the path as shown in Fig. 11. A light beam with electric field parallel to the plane of the paper (H) gets transmitted, and one with electric field perpendicular to the plane of the paper (V) gets reflected. Hence, it is a successful realization of a CNOT gate.
The truth table for this CNOT gate is
\vskip0.1in
\begin{tabular}{c|c|c}
\centering

Control Bit & Target Bit&  	Output\\
H			&	0	 	&		0\\
H			&	1	 	&		1\\
V			&	0	 	&		1\\
V			&	1	 	&		0	
\end{tabular}

\subsubsection*{Toffoli Gate}
In the quantum case a Toffoli gate performs a NOT operation on a target qubit depending on the state of two control qubits. In the classical optical case,
consider the set of two control path cebits 00,01,10,11 representing four light beams, as shown in Fig. 12. The initial state of the beams is
\begin{equation}
|\psi\rangle=\frac{1}{2}\left[|00\rangle|H\rangle+|01\rangle|H\rangle+|10\rangle|H\rangle+|11\rangle|H\rangle\right].
\end{equation}
As only the last beam marked by 11 passes through the half-wave plate HWP, only its polarization gets flipped from H to V, and hence the state changes to 
\begin{equation}
|\psi\rangle^\prime=\frac{1}{2}\left[|00\rangle|H\rangle+|01\rangle|H\rangle+|10\rangle|H\rangle+|11\rangle|V\rangle\right]
\end{equation}
Clearly, the polarization cebits are the target cebits, there are two control path cebits, and only the last target cebit is flipped. Hence, this is a classical realization of a Toffoli gate.

\begin{figure}[ht]
\centering
\hspace*{1.6cm}
\includegraphics[scale=0.7]{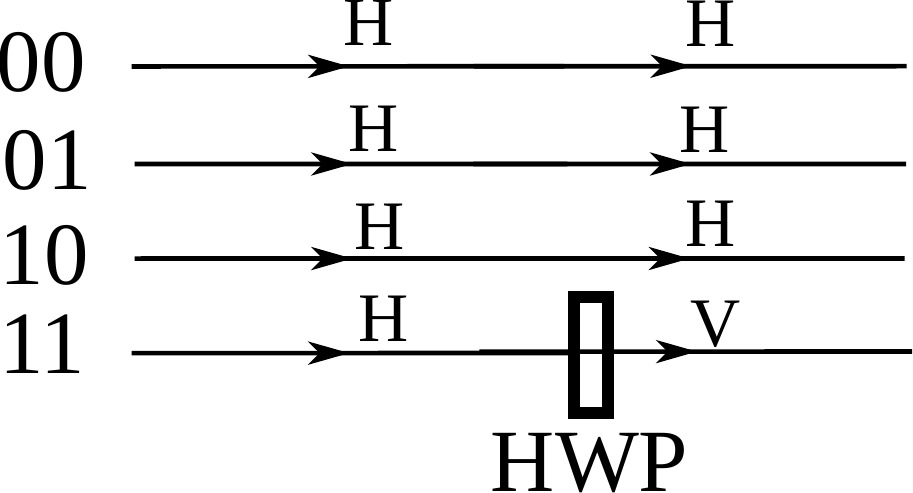} 
\caption{\label{Figure 12}{\footnotesize A Toffoli gate with propagation mode as the control cebit. HWP is a half-wave plate with its fast axis oriented at $45^{0}$ to the vertical.}}
\end{figure} 

\section*{Teleportation}

\begin{figure}[ht]
\includegraphics[scale=1]{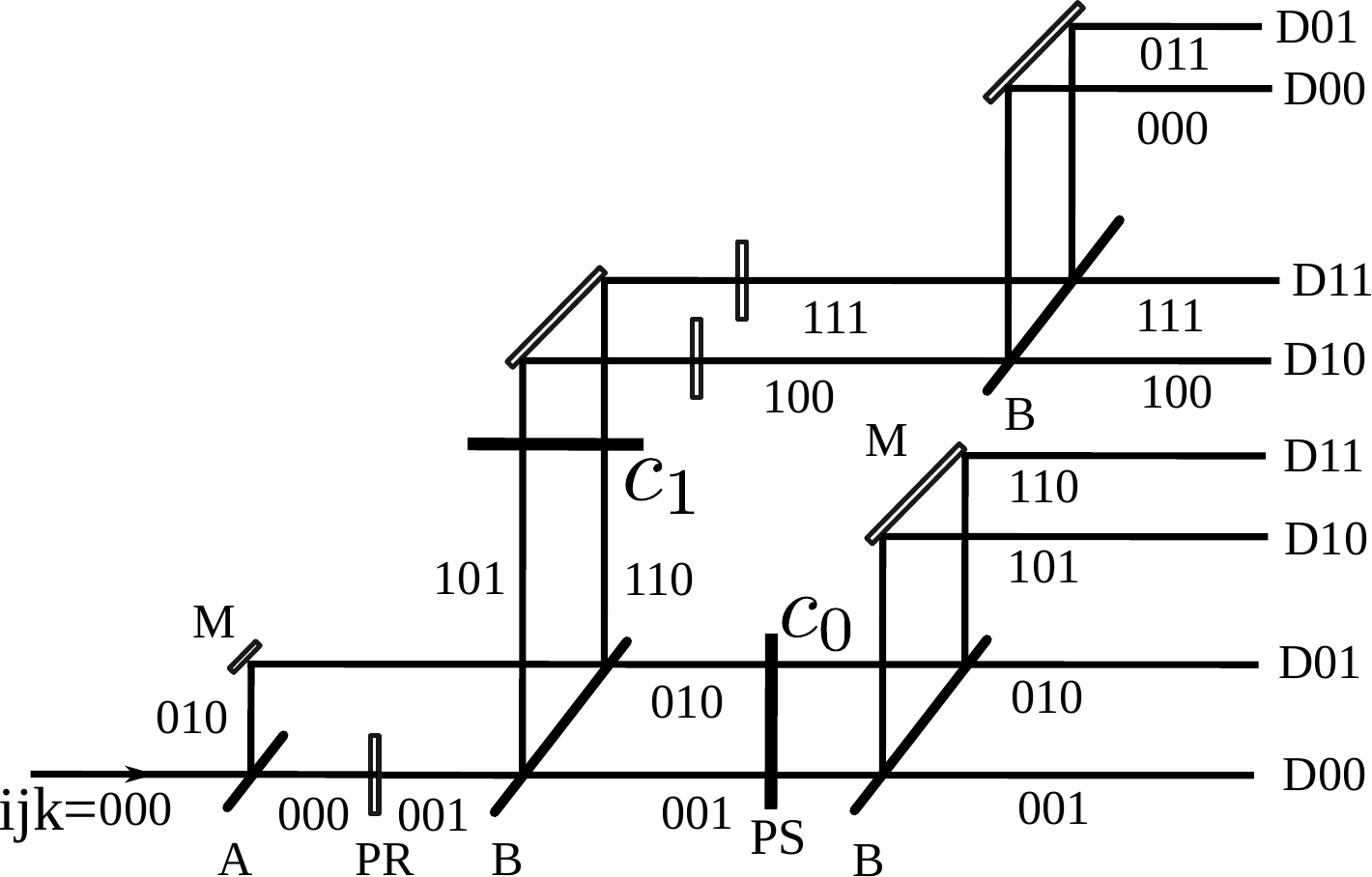} 
\caption{\label{Figure 13}{\footnotesize The first level symmetric beam splitters are marked A and the second level B; the first index i gives information about splitting at B, the second index j gives information about splitting at A, and the third index k gives information regarding polarization, i.e. 0 for horizontal H and 1 for vertical polarization V. PR is a polarization rotator which rotates the polarization from 0 to 1 and vice-versa. PS assigns phases $c_{0}$ and $c_{1}$ to the light paths specified by i, the first index.}}
\end{figure}

In quantum teleportation, typically a sender Alice transmits the quantum state of a qubit to a distant receiver Bob by using dual channels, one classical and the other an entangled (i.e. EPR) channel shared by them. Let us see how an analogous effect using classical optics can be realized. This example will involve no nonlocality.
Let the initial state of light be denoted by $|000\rangle$, the first two indices specifying its location and the third index its polarization (Fig. 13). The last index tells us that the light is H polarized and is incident on a first level beam splitter A. Since A operates on the second cebit, it is analogous to a Hadamard gate 

$$ H=\dfrac{1}{\sqrt{2}}\left(\begin{array}{cc}
				1	&	1\\
				-1	&	1																			
				\end{array}\right),$$
and 				
\begin{equation}
|000\rangle\longrightarrow |0\rangle\left[\frac{|0\rangle+ |1\rangle}{\sqrt{2}}\right]|0\rangle.
\end{equation}
The overall unitary operation on the initial state is thus $U=I\otimes H\otimes I$.

Since the polarization rotator PR rotates the polarization in the $|00\rangle$ path from $|0\rangle$(horizontal)$\rightarrow$ $|1\rangle$(vertical), we have
\begin{equation}
\frac{1}{\sqrt{2}}\left[|000\rangle+ |010\rangle\right]\longrightarrow |0\rangle\left[\frac{|01\rangle + |10\rangle}{\sqrt{2}}\right].
\end{equation}
Notice that the last two cebits of this state are entangled, the entanglement being between the path and polarization degrees of freedom of the beam. This is analogous to creating a quantum entangled EPR pair to be shared by Alice and Bob. 
The action of a B beam splitter which operates on the first cebit is also analogous to a Hadamard gate, and hence 
\begin{equation}
|0\rangle\left[\frac{|01\rangle + |10\rangle}{\sqrt{2}}\right]\rightarrow \left[\frac{|0\rangle+ |1\rangle}{\sqrt{2}}\right]\left[\frac{|01\rangle + |10\rangle}{\sqrt{2}}\right].
\end{equation}
The overall unitary operation on the state is thus $U=H\otimes I\otimes I$.
This state can be rewritten in the form
\begin{equation}
|\Psi\rangle=\frac{1}{2}\left([|00\rangle |1\rangle + |01\rangle |0\rangle]+[|10\rangle |1\rangle + |11\rangle |0\rangle]\right)
\end{equation}
To create the cebit to be teleported, phase shifters PS are placed in the pair of paths ($|00\rangle$, $|01\rangle$) and ($|10\rangle$, $|11\rangle$). They introduce phase lags $c_{0}$ and $c_{1}$ in these pairs of paths. Therefore the state changes to
\begin{equation}
|\Psi\rangle^* = \frac{1}{\sqrt{2}}|0\rangle \left[c_{0}(|10\rangle+ |01\rangle)\right]+\frac{1}{\sqrt{2}}|1\rangle\left[c_{1}(|10\rangle+ |01\rangle)\right]
\end{equation}
Let the values of these phases attached to the paths be arbitrary. The aim is to transfer this phase information from the path cebit to the polarization cebit. Since path and polarization belong to disjoint Hilbert spaces, this is analogous to quantum teleportation between disjoint coordinate spaces.

Next, the action of the polarization rotators PR on the states ($|110\rangle$, $|101\rangle$) is to change the last cebit from 0 to 1 and vice versa. Therefore, the state evolves to
\begin{equation}
|\Psi\rangle^{**} =\frac{1}{\sqrt{2}}|0\rangle \left[c_{0}(|10\rangle+ |01\rangle)\right]+\frac{1}{\sqrt{2}}|1\rangle\left[c_{1}(|00\rangle+ |11\rangle)\right].
\end{equation}
Finally, the action of the B beam splitter on the pair of paths ($|00\rangle$, $|01\rangle$) and ($|10\rangle$, $|11\rangle$) is to produce the final state
\begin{equation}
|\Psi\rangle_f =\frac{1}{2}\left[|0\rangle + |1\rangle\right]\left[c_{0}(|10\rangle+ |01\rangle)\right]+\frac{1}{2}\left[|1\rangle - |0\rangle\right]\left[c_{1}(|00\rangle+ |11\rangle)\right]
\end{equation}
This can be written in the more illuminating form
\begin{eqnarray}
|\Psi\rangle=\frac{1}{2}\left(|00\rangle\left[c_{0}|1\rangle -c_{1}|0\rangle\right]+|01\rangle\left[c_{0}|0\rangle -c_{1}|1\rangle\right]\right)\nonumber\\+\frac{1}{2}\left(|10\rangle\left[c_{0}|1\rangle +c_{1}|0\rangle\right]+|11\rangle\left[c_{0}|0\rangle +c_{1}|1\rangle\right]\right)
\end{eqnarray}
which shows that the arrangement is successful in transfering the unknown phase information ($c_{0}$, $c_{1}$) from the path space to the disjoint polarization space. The phase information can be read off by measuring the polarizations in the paths $\lbrace 00,01,10,11\rbrace$.
 
\section{Concluding Remarks}
We have shown how useful and instructive it is to use the Hilbert space structure of classical optics (electrodynamics), and how it leads to novel aspects of classical light. Before concluding we would like to point out another potentially useful area of application of the Hilbert space structure of classical field theories, though it lies strictly outside the realm of hard core physics. Quantum cognition is an emergent area of cognitive modeling that is novel and has many advantages over theories based on classical logic and classical probability theory \cite{aerts, pothos}. The main motivation for searching beyond the limits of classical logic and classical probability theory as a basis for modeling cognitive processes is empirical evidence such as the `Guppy effect' \cite{osh}. The overextension and underextension of membership weights of items that it implies cannot be explained by any classical model \cite{tversky,hampton,gabora}. The central aspect of quantum mechanics that the proponents of `quantum cognition' exploit to overcome the inadequacies of classical theory is essentially its Hilbert space structure that naturally incorporates superposition, interference, incompatibility, contextuality and order, and entanglement. The claim is that this structure lies at the origin of specific effects in cognition related to the way in which concepts and their combinations carry and influence thir meaning. The example given by Osheron and Smith is the pair of concepts {\em Pet} and {\em Fish} and their conjunction {\em Pet-and-Fish} or their disjunction {\em Pet-or-Fish}. They observed that while an exemplar or item such as Guppy was a very typical example of Pet-Fish, it was neither a very typical example of Pet nor of Fish. This psychological behaviour is unexpected in classical theory. What is shown is that quantum modeling gives a more natural and satisfactory account of such psychological/cognitive processes, and reduces to classical modeling in special circumstances \cite{sozzo}. Two other examples where classical modeling is found to be inadequate are: (1) a person who is neither happy nor unhappy can be in a superposition of happy and unhappy states which is not permitted in classical modeling; (2) an empirical answer to two incompatible questions such as whether Linda is both a feminist and a bank teller has been found to be inconsistent with the predictions of classical probability theory.

However, the fundamental question that has not been addressed by the proponents of quantum cognition is: how does this typically quantum structure survive in macroscopic systems like the brain which are thermodynamically open? It is well known that quantum systems are fragile and subject to rapid decoherence that quenches their quantum properties. We have shown that Hilbert space structure is not exclusive to quantum mechanics, it is an inherent property of paradigmatic classical theories such as Maxwellian electrodynamics. The fact that coherence and entanglement are robust in classical field theories makes them eminently more suitable for cognitive modelling than quantum theory.
\vskip 0.1in
{\flushleft {\em Note added}

Our attention has just been drawn to a paper by X-F Qian and J. H. Eberly (arXiv:1307.3772 [quant-ph] 14 July 2013) that deals with entanglement in classical thermal light and Bell analyses. This is a sequel to their earlier paper in which they reformulated polarization theory as entanglement analysis \cite{eberly}. The present review deals with deterministic light.

\section{Acknowledgement}
PG thanks the National Academy of Sciences, India for the award of a Senior Scientist Platinum Jubilee Fellowship which allowed this work to be undertaken.

\end{document}